\documentclass[prd,aps,eqsecnum,amsmath,floatfix,nofootinbib,preprint,tightenlines]{revtex4-1}

\usepackage{latexsym}
\usepackage{graphicx}
\usepackage{multirow}
\usepackage[dvipsnames]{xcolor}

\usepackage{hyperref}

\def\floatcaption#1#2{ \caption{#2 \label{#1}} }

\def\bibi{\bibitem}
\def\ttl#1{{\it #1}}

\let\slo=\o                     % slashed o (Scandinavian)
\def\a{\alpha}
\def\b{\beta}
\def\c{\chi}
\def\d{\delta}
\def\g{\gamma}
\def\h{\eta}

\def\j{\psi}

\def\l{\lambda}
\def\m{\mu}
\def\n{\nu}
\def\o{\omega}
\def\p{\pi}                     % Also, \varpi
\def\r{\rho}                    %       \varrho
\def\s{\sigma}                  %       \varsigma
\def\t{\tau}

\def\D{\Delta}

\def\G{\Gamma}

\def\L{\Lambda}

\def\S{\Sigma}

\def\cl{{\cal L}}
\def\cm{{\cal M}}

\def\co{{\cal O}}

\def\cbo{{\,\raise-.15ex\Sc [\,}}                       % curly "
\def\ltap{\raisebox{-.4ex}{\rlap{$\sim$}} \raisebox{.4ex}{$<$}}   % < or ~
\def\svev#1{\left\langle #1\right\rangle}       % variable < >
\def\ddt#1{{\buildrel {\hbox{\LARGE .\kern-2pt.}} \over {#1}}}% double dot-over
\def\ie{\mbox{\it i.e.}}

\def\etc{\mbox{\it etc.}}

\def\tr{{\rm tr}\,}

\def\half{{1\over 2}}

\def\seef{{\it cf.\  }}

\def\tm{\tilde\m}
\def\tb{\tilde\b}
\def\tc{\tilde{c}}
\def\tV{\tilde{V}}

\def\hB{\hat{B}}
\def\hm{\hat{\m}}
\def\td{\tilde{d}}

\def\hf{\hat{f}}

\def\bj{\overline{\j}}

\def\LUV{\L_{UV}}

\begin{document}

\begin{boldmath}
\begin{center}
{\large{\bf
Explorations beyond dilaton chiral perturbation theory\\[4mm]
in the eight-flavor SU(3) gauge theory
}}\\[8mm]
Maarten Golterman$^a$ and Yigal Shamir$^b$\\[8 mm]
$^a$Department of Physics and Astronomy, San Francisco State University,\\
San Francisco, CA 94132, USA\\
$^b$Raymond and Beverly Sackler School of Physics and Astronomy,\\
Tel~Aviv University, 69978, Tel~Aviv, Israel\\[10mm]
\end{center}
\end{boldmath}

\begin{quotation}
We continue our study of spectroscopy data for the SU(3) gauge theory
with eight fundamental fermions, motivated by the effective field theory
framework of dilaton chiral perturbation theory (dChPT).
At leading order dChPT predicts a constant mass anomalous dimension $\g_m$,
consistent with the assumed proximity of an infrared fixed point.
For the relatively large fermion masses simulated by the LatKMI collaboration,
the influence of the infrared fixed point diminishes, and our fits suggest
that $\g_m$ starts running.
Since a complete higher-order analysis is not feasible
with presently available data, we adopt a more phenomenological approach.
We propose a partial extension
to higher orders, which incorporates the running of $\g_m$ into the
tree-level lagrangian.  We find that this extension successfully describes
the full fermion-mass range of the LatKMI data, including the pion
taste splittings which arise from using staggered fermions in the
lattice simulations.
We also investigate a more general class of dilaton potentials
proposed in the literature, using both the LSD and LatKMI data sets,
concluding that these data favor the form predicted by dChPT.
\end{quotation}

%%%%%%%%%%%%%%%%%%%%%%%%%%%
\newpage
\section{\label{intro} Introduction}
%%%%%%%%%%%%%%%%%%%%%%%%%%%
Lattice simulations of the SU(3) gauge theory with eight
Dirac fermions in the fundamental representation have revealed
the existence of a flavor-singlet scalar particle, which,
at the fermion masses explored in these simulations, is approximately
degenerate with the pions---the Nambu--Goldstone bosons
associated with chiral symmetry breaking \cite{LSD0,LSD,LSD2,LatKMI}.
A similar light scalar has been found also in the SU(3) gauge theory
with two sextet fermions \cite{sextetconn,sextet1,sextet2,Kutietal,Kutietal20},
or with four light, and six
\cite{LSD10} or eight \cite{BHRWW} heavy fundamental fermions.\footnote{%
 For reviews of lattice work, see Refs.~\cite{DeGrandreview,NP,Pica,Benlat17,Drach}.
}

The existence of a light flavor-singlet scalar particle roughly degenerate
with the pions means that, besides the pions,
any effective field theory (EFT) description of the low-energy behavior
has to include a field that represents this scalar particle.
Here, our starting point is dilaton chiral perturbation theory (dChPT),
an EFT in which the lightness of the scalar particle is assumed to arise
from approximate scale invariance of the underlying theory
in the infrared \cite{PP,latt16,gammay,largemass,BGKSS}.\footnote{
  For early work, and for other low-energy approaches, see
  Refs.~\cite{ApB,BLL,GGS,CM,MY,HLS,AIP1,AIP2,CT,LSDsigma,AIP3}.
}
Increasing the number of (massless) fermionic degrees of freedom
will eventually take the theory into the conformal window,
where the non-abelian gauge theory is still asymptotically free,
but develops an infrared fixed point (IRFP).  The idea is that,
with eight flavors, the SU(3) gauge theory is still outside
the conformal window, but close enough to the conformal sill---the
number of flavors where the IRFP first develops---that the breaking of
scale invariance in the infrared is governed by the proximity of the IRFP.
The key assumption is then that the distance to the conformal sill
can be treated as a small parameter, in which a systematic power counting
can be developed.  The scalar particle, which we will refer to as
the dilaton, is interpreted as a pseudo Nambu--Goldstone boson (pNGB)
for the approximate scale symmetry \cite{PP}.  The mass of the dilaton is
controlled by this small parameter, just as the fermion mass leads to
a parametrically small pion mass.
Since the fermion mass breaks scale invariance too,
the dilaton mass will also depend on the fermion mass.

In a previous paper \cite{GNS} we applied leading-order (LO) dChPT
to numerical data for the eight-flavor SU(3) gauge theory
produced in lattice simulations by the LSD collaboration \cite{LSD2}.
We showed that, over the fermion mass range in these simulations,
LO dChPT successfully describes the pNGB sector of the theory,
including the dilaton.  In Ref.~\cite{LSD2} staggered fermions were used,
which exhibit taste splittings---a lattice artifact mass splitting
of the pion multiplet caused by a partial breaking of the
flavor symmetry group in the staggered fermion formulation.\footnote{%
  For reviews, see Refs.~\cite{MILC,MGLH}.
}
We showed that dChPT explains the pattern of taste splittings
in the pion sector observed in Ref.~\cite{LSD2} as a function of the fermion mass.
The vacuum expectation value of the dilaton field depends
on the fermion mass already in LO, leading to a fermion-mass dependence
of pNGB decay constants and masses that is qualitatively different from QCD.
This includes the taste splittings, which are also qualitatively different
from the pattern seen in QCD with staggered fermions.

Given this success, our goal in this paper is to investigate
whether dChPT can also be applied to the other major lattice study
of the eight-flavor SU(3) gauge theory, by the LatKMI collaboration
\cite{LatKMI}.\footnote{%
  We will often shorten ``LatKMI'' to just ``KMI.''
}
This study also used staggered fermions, and presented extensive spectroscopy
data for the pNGB sector, including taste splittings.
The KMI simulations were done at larger fermion masses than those of
LSD.  Even if dChPT is the correct EFT, the question arises whether
one can fit the KMI data using LO dChPT, or, alternatively,
whether higher orders in the EFT expansion would be needed.
Indeed, unlike for the LSD data \cite{LSD2},
we found that LO dChPT does not quantitatively describe the KMI data
over the full fermion mass range, as will be discussed in detail in this paper.

For the LSD data, we found that as the fermion mass is varied,
hadron masses and decay constants
respond with an approximate hyperscaling behavior \cite{largemass}.
As the fermion mass increases,
the theory is drawn further away from the influence of the IRFP at the
nearby conformal sill.
Once the fermion mass becomes large enough, we expect
that the running of the coupling will become noticeable, and thus also
the running of the mass anomalous dimension $\g_m$.\footnote{%
  For an early study of $\g_m$ in the $N_f=8$ theory, see Ref.~\cite{Annaetal}.
}
In dChPT, at leading order, the mass anomalous dimension is constant,
$\g_m=\g_*$, where $\g_*$ is the mass anomalous dimension at the nearby IRFP.
dChPT allows for a non-constant $\g_m$, but the power counting
underlying dChPT accommodates corrections to a constant $\g_m$
only through higher orders. In order to systematically
compare dChPT with the KMI data, we would thus have to consider dChPT to
next-to-leading order (NLO) or beyond.  However,  the
relatively large number of additional parameters that would be needed
already at NLO, and limitations of the presently available
lattice data, to be discussed below,
prevent us from attempting a complete NLO fit.

Instead, we will take a more phenomenological approach,
based on the following observation.  The salient
difference between the KMI and LSD data appears to be that
a constant $\g_m$ cannot account for the full range of (larger)
fermion masses explored in the KMI data.
We will thus extend LO dChPT by only including
higher-order effects that are directly related to $\g_m$; we will refer to this
extension as $\g$-dChPT.    This makes our approach not systematic,
since most NLO and higher-order effects are left out.  Strictly speaking,
$\g$-dChPT should thus be viewed as a model approach.

In order for LO dChPT to accommodate a varying $\g_m$, we will modify
the mass-dependent part of the potential, as described in detail
in Sec.~\ref{gammam}.  This raises the question of
what happens if one also considers a generalization of the dilaton part
of the potential.  A class of potentials depending on a new parameter $\D$,
generalizing the dilaton potential of dChPT,  has been proposed before
\cite{GGS,CM,AIP1,AIP3}, and we will refer to this different extension
of LO dChPT as $\D$-dChPT.  One recovers LO dChPT,
including its dilaton potential, by taking $\D\to 4$.
It is interesting to also confront $\D$-dChPT with the data.
We will revisit the analysis of the LSD data using $\D$-dChPT
by Ref.~\cite{AIP3}, and extend this investigation to
the KMI data.   Despite claims in the literature \cite{AIP3},
$\D$-dChPT takes us outside the systematic power counting of dChPT,
and should thus be considered as a more phenomenological
approach to the low-energy behavior of the $N_f=8$ theory.

This paper is organized as follows.   In Sec.~\ref{gammam} we introduce
$\g$-dChPT, in which LO dChPT is extended to accommodate a varying $\g_m$.
In Sec.~\ref{KMI} we first present our evidence that $\g_m$,
as well as other LO parameters, are changing
over the KMI mass range in a fit to LO dChPT.  We then
apply $\g$-dChPT to the pNGB sector of the KMI data.
We find that a rather simple model for a varying $\g_m$ provides
good fits of the KMI data, including taste splittings.
In Sec.~\ref{potential} we consider the generalized class
of dilaton potentials, reviewing the application of $\D$-dChPT to the LSD
data, and applying it to the KMI data.
Combining these results provides some evidence that the dilaton potential
of LO dChPT is preferred by the data, \ie,
that the preferred value in $\D$-dChPT is close to $\D=4$.
Finally, Sec.~\ref{conclusion} contains our conclusions.
In App.~\ref{massindep} we elaborate on the choice of a mass-independent
scale setting prescription.  In App.~\ref{power counting} we investigate
the claim of Ref.~\cite{AIP3} that $\D$-dChPT admits a systematic power counting
for any value of $\D$, and show that this claim is incorrect.

%%%%%%%%%%%%%%%%%%%%%%%%%%%
%\newpage
\begin{boldmath}
\section{\label{gammam} Dilaton C\lowercase{h}PT and $\g_m$}
\end{boldmath}
%%%%%%%%%%%%%%%%%%%%%%%%%%%
In Sec.~\ref{tree}, we begin with a summary of LO dChPT.
This is the EFT that was applied to the LSD data in Ref.~\cite{GNS}.
In Sec.~\ref{hypers} we revisit the physics of hyperscaling,
and its manifestation in LO dChPT.
This leads us in Sec.~\ref{vary} to introduce $\g$-dChPT,
where we generalize the low-energy lagrangian to accommodate
a non-constant mass anomalous dimension.
We emphasize that this extension takes us outside the strict EFT framework.
In Sec.~\ref{hadrons} we present the
hadronic quantities to be fit to the KMI data of Ref.~\cite{LatKMI} in the
rest of this paper.

%%%%%%%%%%%%%%%%%%%%%%%%%%%
%\newpage
\vspace{3ex}
\subsection{\label{tree} dChPT at lowest order}
%%%%%%%%%%%%%%%%%%%%%%%%%%%
The euclidean LO lagrangian for dChPT is given
by
\begin{equation}
\label{LOlag}
\cl=\frac{1}{2}f_\t^2 e^{2\t}\partial_\m\t\partial_\m\t+
\frac{1}{4}f_\p^2 e^{2\t}\,\tr(\partial_\m\S^\dagger\partial_\m\S)+\cl_m(\t,\S)
+\cl_d(\t)\ .
\end{equation}
The potential terms are
\begin{subequations}
\label{potentials}
\begin{eqnarray}
\cl_d(\t)&=&f_\t^2B_\t e^{4\t}\left(c_0+c_1 \t\right)\ ,
\label{potentialsa}\\
\cl_m(\t,\S)&=&-\frac{1}{2}f_\p^2B_\p m e^{(3-\g_*)\t}\,\tr(\S+\S^\dagger)\ .
\label{potentialsb}
\end{eqnarray}
\end{subequations}
Here $\S$ is the usual non-linear field describing the pion multiplet,
while $\t$ is the dilaton effective field.
$\cl$ depends on the low-energy constants (LECs)
$f_\t$, $f_\p$, $B_\p$, $B_\t$, $\g_*$, $c_0$ and $c_1$.

We define the theory in the Veneziano limit \cite{VZlimit},
in which $N\equiv N_c\propto N_f$ is taken to infinity keeping the
ratio $n_f=N_f/N_c$ fixed,
with $N_f$ the number of fundamental-representation flavors and
$N_c$  the number of colors.  The power counting is \cite{PP}
\begin{equation}
\label{pc}
p^2\sim m\sim n_f-n_f^* \sim 1/N\ .
\end{equation}
The relation $p^2 \sim m$ defines the power counting
of ordinary ChPT.\footnote{%
  The dimensionful quantities, $p^2$ and $m$, are measured in units
  of the dynamically generated infrared scale of the massless theory.
}
The small parameter controlling the hard breaking
of scale invariance is $n_f-n_f^*$, where $n_f^*$ is the limiting value of $n_f$
for the theory at the conformal sill: the boundary between the regime
where the massless theory undergoes chiral symmetry breaking, and the regime
where this theory is conformal in the infrared,
\ie, where the gauge coupling $g$ runs into an infrared fixed point $g_*$.

Invoking the proximity of the sill of the conformal window,
we assume that the $\b$ function is small at the chiral symmetry breaking
scale, and that the corresponding value of $g$ is close to $g_*$.
We can then expand the mass anomalous dimension $\g(g)$ in powers of $n_f-n_f^*$
around $\g_*=\g(g_*)$, the mass anomalous dimension at the infrared fixed point
at the conformal sill. For a detailed discussion of the construction
of the LO lagrangian, and the underlying power counting,
see Refs.~\cite{PP,largemass}.

In the dilaton potential~(\ref{potentialsa}), $c_0$ is $\co(1)$, while
$c_1$ is proportional to the small expansion parameter $n_f-n_f^*$.\footnote{%
  For a few more details about the power counting, see App.~\ref{power counting}.
}
For $m=0$, we shift the $\t$ field to $\t+v_0$,
with $v_0=\svev{\t}\big|_{m=0}$ (before the shift).
After the shift, the dilaton expectation value
$v(m)=\svev{\t}$ vanishes in the massless theory.  Defining
\begin{subequations}
\label{shiftlec}
\begin{eqnarray}
\label{shiftleca}
\hf_{\p,\t}&=&e^{v_0}f_{\p,\t}\ ,
\\
\label{shiftlecb}
\hB_\t&=& e^{2v_0}B_\t\ ,
\\
\label{shiftlecc}
\hB_\p&=& e^{(1-\g_*)v_0}B_\p\ ,
\end{eqnarray}
\end{subequations}
the lagrangian becomes
\begin{equation}
\label{LOlagredef}
\cl=\frac{1}{2}\hf_\t^2 e^{2\t}\partial_\m\t\partial_\m\t+
\frac{1}{4}\hf_\p^2 e^{2\t}\,\tr(\partial_\m\S^\dagger\partial_\m\S)+\cl_m(\t,\S)
+\cl_d(\t)\ ,
\end{equation}
with
\begin{subequations}
\label{hatpot}
\begin{eqnarray}
\cl_d(\t)&=& \hf_\t^2\hB_\t e^{4\t}\, V_d(\t)\ ,
\label{hatpota}\\
V_d(\t) &=& c_1 \left(\t-\frac{1}{4}\right) \ ,
\label{Vd}\\
\cl_m(\t,\S)&=&-\frac{1}{2}\hf_\p^2\hB_\p m e^{(3-\g_*)\t}\,\tr(\S+\S^\dagger)\ .
\label{hatpotb}
\end{eqnarray}
\end{subequations}
The shift sets $c_0=-c_1/4$, and now the whole LO lagrangian
is $\co(p^2)$ in the power counting~(\ref{pc}).
We will assume $c_1>0$, so that the potential $\cl_d+\cl_m$ is
bounded from below.

Assuming $m\ge 0$, the potential is minimized by $\S=1$.
The dilaton expectation value $v=v(m)$ solves the saddle-point equation
\begin{equation}
\label{saddle}
\frac{(3-\g_*)m}{4c_1\cm}=v\,e^{(1+\g_*)v}\ ,\qquad
\cm=\frac{\hf_\t^2\hB_\t}{\hf_\p^2\hB_\p N_f}\ .
\end{equation}
The solution is positive, and monotonically increasing with $m$.
The spectroscopy data we considered in Ref.~\cite{GNS} can then be expressed
as functions of $m$,
\begin{subequations}
\label{quant}
\begin{eqnarray}
\frac{M_\p^2}{F_\p^2}&=&\frac{1}{d_1}\,v(m)\equiv h(m)\ ,\label{quanta}\\
F_\p&=&\hf_\p\,e^{v(m)}\label{quantb}\\
&=&\left(\frac{d_0 m}{h(m)}\right)^{\frac{1}{1+\g_*}}\ ,\label{quantc}\\
\frac{M_\t^2}{F_\p^2}&=&d_3\left(1+(1+\g_*)\,d_1 h(m)\right)\ .\label{quantd}
\end{eqnarray}
\end{subequations}
Explicitly,
\begin{equation}
\label{x}
h(m)=\frac{1}{(1+\g_*)d_1}\,W_0\left(\frac{(1+\g_*)d_1}{d_2}\,m\right)\ ,
\end{equation}
where $W_0$ is the Lambert $W$-function.
The parameters $d_{0,1,2,3}$ are defined
in terms of the LECs of the tree-level lagrangian,
\begin{equation}
\label{ds}
d_0=\frac{2\hB_\p}{\hf_\p^{1-\g_*}}\ ,\quad
d_1=\frac{(3-\g_*)\hf_\p^2}{8\hB_\p c_1\cm}\ ,\quad
d_2=\frac{\hf_\p^2}{2\hB_\p }\ ,\quad
d_3 =\frac{4 c_1 \hB_\t}{\hf_\p^2} \ .
\end{equation}

In Ref.~\cite{GNS} we applied LO dChPT, as summarized above, to the
LSD data \cite{LSD2}.  The key assumptions underlying this analysis were:
(a) the $N_f=8$, $N_c=3$ theory undergoes chiral symmetry breaking;
(b) for the LSD mass range, the $\b$ function is small enough
that the dChPT power counting is applicable.
The results of our analysis corroborated these assumptions.

%%%%%%%%%%%%%%%%%%%%%%%%%%%
%\newpage
\subsection{\label{hypers} Hyperscaling}
%%%%%%%%%%%%%%%%%%%%%%%%%%%
Consider momentarily a mass-deformed infrared conformal theory.
We can probe the theory over a range of scales where $g$
is so close to the infrared fixed-point $g_*$ that all effects
of its running can be neglected.  The breaking of scale invariance
is then driven entirely by the input bare fermion mass $m_0$.
Under these circumstances,
any hadronic mass $M$ follows a simple hyperscaling law,
\begin{equation}
\label{hs}
\frac{M}{\L_{\rm UV}}\sim \left(\frac{m_0}{\L_{\rm UV}}\right)^{\frac{1}{1+\g_*}}\ .
\end{equation}
Here $\L_{\rm UV}$ is an ultraviolet scale for which the approximation
$\g_m(\m)=\g_*$ is valid for any $\m\le \LUV$,
and $m_0=m(\LUV)$, where $m(\m)$ is the running renormalized mass.
Hyperscaling is based on the following simple observations:
\begin{enumerate}
\item The renormalized mass, $m=m(\mu)$, runs as dictated by
its anomalous dimension.  By contrast, the renormalized coupling
has attained its fixed-point value $g_*$ (up to negligible corrections),
hence also the mass anomalous dimension has a fixed value $\g_*=\g_m(g_*)$.
\item No physical scale is generated dynamically in the massless theory.
When the fermion mass is non-zero, the induced physical scale $M$ is set by
the condition $M \sim m(M)$.
\end{enumerate}
Indeed, starting from the solution for $m(\m)$ for a constant
mass anomalous dimension,
\begin{equation}
\label{runmsol}
\frac{m(\m)}{m_0} = \left(\frac{\m}{\LUV}\right)^{-\g_*}\ ,
\end{equation}
the hyperscaling law~(\ref{hs}) immediately follows by postulating
that the typical hadron mass $M$ satisfies $M \sim m(\m)$ for $\m=M$.
For any $\g_* > 0$, the existence of the physical scale $M$ is guaranteed
if $m_0\ll \LUV$.  Starting from $m(\m)=m_0 \ll \m$ at $\m=\LUV$,
$m(\m)$ keeps increasing as $\m$ is decreased, until eventually
the equality $M=m(M)$ is reached.

Returning to dChPT, in
Ref.~\cite{GNS} we found that the LSD data is in the ``large-mass'' regime
\cite{largemass}, where
\begin{equation}
\label{LM}
|n_f-n_f^*| \sim c_1 \ll \frac{m_0}{\cm} \ ,
\end{equation}
for all (bare) masses. As follows from the previous subsection,\footnote{%
  See also App.~\ref{power counting}.}
in LO dChPT, $c_1$ encodes
the magnitude of the $\b$ function at the chiral symmetry breaking scale.
The large-mass regime is thus an approximate hyperscaling regime,
where the input fermion mass dominates the breaking of scale invariance.
Indeed, in Ref.~\cite{largemass} we showed that the leading mass dependence
predicted by LO dChPT in the large-mass regime is the hyperscaling
relation~(\ref{hs}), for all hadronic masses and decay constants.
We also calculated corrections to this relation,
which are present in dChPT already at LO, because the $\b$ function
at the chiral symmetry breaking scale, hence $c_1$,
is (by assumption) parametrically small,
but not vanishingly small as in a mass-deformed infrared conformal theory.
Moreover, we showed that as long as
\begin{equation}
\label{LMexpand}
  |n_f-n_f^*| \log \left(\frac{m_0}{|n_f-n_f^*| \cm}\right) \ll 1 \ ,
\end{equation}
dChPT provide a systematic expansion, even though $m_0/\cm$ can be large.
By Eq.~(\ref{saddle}), $\cm$ is constructed from LECs
which can be defined in the chiral limit.
It is a striking difference between ordinary ChPT and dChPT that,
because of the nearby IRFP,
in dChPT a systematic low-energy expansion exists even if the fermion mass
is not small relative to the infrared scale of the massless theory, so long as
inequality~(\ref{LMexpand}) holds.

The fermion mass range explored in the KMI data is higher than
in the LSD data.  The comparison can be made, for example, in units of $t_0$,
see Fig.~5 of Ref.~\cite{LSD}.  We will return to the comparison between
the LSD and KMI data, and its limitations, in Sec.~\ref{KMIdisc} below.
As mentioned in the introduction, when we increase the input
fermion mass the influence of the IRFP diminishes.
Eventually, we will reach energy scales where
the running of the coupling picks up,\footnote{%
  At extremely high energy scales perturbation theory will eventually
  take over, and the $\b$ function will tend to zero as dictated by
  asymptotic freedom.
}
and, as a result, so does
the running of the mass anomalous dimension.
In the next subsection, guided by this consideration,
we will develop a generalized notion of hyperscaling,
which is founded on the same principles as above, except that the assumption
of a constant mass anomalous dimension is relaxed.
This will lead to the framework of $\g$-dChPT,
where LO dChPT is extended to accommodate a varying mass anomalous dimension.
We stress that the power counting of dChPT
allows for corrections to a constant $\g_m$, but only via higher-order terms
in the expansion in $n_f-n_f^*$.  In seeking an extension of LO dChPT
that accommodates a varying $\g_m$ we are thus asking for a partial resummation
of these higher-order terms, under the assumption that these are
the dominant higher-order corrections.

We conclude this subsection with a technical comment.
The hyperscaling law~(\ref{hs}) can be rewritten as
\begin{equation}
\label{hsrew}
\frac{m_0}{M}\sim\left(\frac{m_0}{\L_{\rm UV}}\right)^{\frac{\g_*}{1+\g_*}}\ .
\end{equation}
It follows that the fermion mass $m_0$ is always much smaller than
any hadronic mass $M$ (as long as $m_0\ll\L_{\rm UV}$),
and the same is true for the decay constants $F_\p$ and $F_\t$.
Moreover, in Ref.~\cite{largemass} we showed that this conclusion
extends to $n_f<n_f^*$, below the conformal window, and that it applies
also to the masses of the pNGBs, $M_\p$ and $M_\t$.
We will assume that the ratio $m_0/M$ remains small also when
the simple hyperscaling relations, Eqs.~(\ref{hs}) and~(\ref{hsrew}),
are generalized to account for the running of $\g_m$.
Indeed, for the LSD data, $m_0/M_\p$ ranges between $0.015$ and $0.04$,
while for the KMI data it ranges between $0.07$ and $0.17$.
Since $m_0/M_\p\ll 1$, this allows us to use a {\em mass-independent}
renormalization scheme.\footnote{%
  The $\b$ and $\g$ functions in a mass-dependent scheme can be expanded
  in powers of $m_0/M$, and the first term in this expansion yields
  a mass-independent scheme that is a good approximation if $m_0/M\ll 1$.
}
As we will see below, this greatly simplifies our considerations.

%%%%%%%%%%%%%%%%%%%%%%%%%%%
%\newpage
\begin{boldmath}
\subsection{\label{vary} Varying $\g_m$ and $\g$-dChPT}
\end{boldmath}
%%%%%%%%%%%%%%%%%%%%%%%%%%%
We will now proceed to develop the extension of LO dChPT allowing
for a scale-dependent $\g_m$.
The RG equation governing the dependence of the renormalized
mass $m$ on the renormalization scale $\m$
is closely related to the behavior of the renormalized mass
under scale transformations.   In order to relate the two,
we first review how a scale is introduced into the bare theory;
we will do this using dimensional regularization.
For more details, we refer to Ref.~\cite{gammay}.
We regulate the action of the microscopic theory as
\begin{equation}
\label{dimreg}
S=\int d^dx\, \m_0^{d-4} \cl(x)\ ,
\end{equation}
where $\cl$ is the bare lagrangian, and $d$ is the number of dimensions.
With the factor $\m_0^{d-4}$, the bare action $S$ is invariant
under scale transformations
if we promote the bare parameters $\m_0$ and $m_0$ to spurions.
The scale transformation rules are
\begin{subequations}
\label{bare}
\begin{eqnarray}
m_0&\to&\l m_0\ ,\label{barea}\\
\m_0&\to&\l\m_0\ ,\label{bareb}\\
A_\m(x) &\to& \l\, A_\m(\l x) \ , \label{barec}\\
\j(x) &\to& \l^{3/2}\,\j(\l x)\ ,\label{bared}
\end{eqnarray}
\end{subequations}
where $A_\m$ is the bare gauge field and $\j$ the bare fermion field.

The function $\g_m$, defined by the RG equation
\begin{equation}
\label{gamma}
\frac{\m}{m}\frac{dm}{d\m}=-\g_m\ ,
\end{equation}
describes the response of the renormalized mass $m$ to a change
of the renormalization scale $\m$.
In a mass-independent scheme, all renormalization factors depend
on the scales $\m$ and $\m_0$ only through their ratio, $\m/\m_0$.
Hence,
\begin{equation}
\label{gammag}
  \g_m = \g_m(g(\m/\m_0)) \ ,
\end{equation}
where $g=g(\m/\m_0)$ is the running coupling.   From now on,
we will write $\g_m(\m/\m_0)$ for $\g_m(g(\m/\m_0))$, with slight
abuse of notation.
We choose $\m$ not to transform under scale transformations:
the transformation~(\ref{bare}) describes a rescaling
of all the dimensionful bare quantities relative to
a fixed renormalization scale.

Once $\g_m$ is known we can express $m(\m)$,
the renormalized mass at an arbitrary renormalization scale $\m$,
in terms of the bare mass, $m_0=m(\m_0)$,
by integrating Eq.~(\ref{gamma}) between $\m_0$ and $\m$.
Introducing the formal solutions
\begin{equation}
\label{RGsol}
E^\pm(\m/\m_0)=e^{\pm\int_0^{\log{\m/\m_0}}dt\,\g_m(e^t)}\ .
\end{equation}
of the RG equations
\begin{equation}
\label{RG}
\m\frac{dE^\pm}{d\m}=\pm\g_m(\m/\m_0)E^\pm\ ,
\end{equation}
one has
\begin{equation}
\label{massren}
m(\m)=E^-(\m/\m_0)\,m_0\ .
\end{equation}
Using Eq.~(\ref{bare}) for the dependence of the bare parameters $m_0$ and $\m_0$
on the scale transformation parameter $\l$, it
follows that an infinitesimal scale transformation
of the renormalized mass is governed by the differential equation
\cite{gammay}
\begin{equation}
\label{dmdlgen}
  \frac{\partial m(\l;\m)}{\partial \log\l}
  = \left(1+\g_m\left(\frac{\m}{\l\m_0}\right)\right) m(\l;\m) \ ,
\end{equation}
which is solved by
\begin{equation}
\label{mtransf}
m(\l;\m)=\l\, E^-(\m/(\l\m_0))\,m_0\ .
\end{equation}
For constant $\g_m=\g_*$,
Eq.~(\ref{RGsol}) simplifies to
\begin{equation}
\label{Egstar}
E^\pm(\m/\m_0)=\left(\frac{\m}{\m_0}\right)^{\pm\g_*} \ ,
\end{equation}
hence
\begin{equation}
\label{gstar}
m(\l;\m)=\l^{1+\g_*}m(\m)=\l^{1+\g_*}\left(\frac{\m_0}{\m}\right)^{\g_*}m_0\ .
\end{equation}
The second equation explains the origin of the factor $\l^{1+\g_*}$.
A factor $\l$ comes from the transformation of $m_0$, Eq.~(\ref{barea}),
while the remaining factor $\l^{\g_*}$ comes from
the transformation of $\m_0$, Eq.~(\ref{bareb}).  With the transformation rules
of the effective fields
\begin{subequations}
\label{diltransf}
\begin{eqnarray}
\label{diltransfa}
\t(x)&\to&\t(\l x)+\log\l\ ,\\
\label{diltransfb}
\S(x)&\to&\S(\l x)\ ,
\end{eqnarray}
\end{subequations}
it follows that $\cl_m(x)$ in Eq.~(\ref{potentialsb}) transforms
into $\l^4\cl_m(\l x)$, as required for the invariance of the action.

In order to accommodate a non-constant $\g_m$, we
replace $\cl_m$ of Eq.~(\ref{potentialsb}) by
\begin{equation}
\label{calBpion}
  \cl_m = -\half f_\p^2 e^{3\t} E^-(e^{\t}f_\p/\m_0) B_\p(\m/\m_0) m(\m/\m_0)\,
  \tr(\S+\S^\dagger) \ .
\end{equation}
Let us derive the transformation properties of this lagrangian.
The combination $B_\p(\m/\m_0) m(\m/\m_0)$ is by assumption
RG invariant, and we can write $B_\p(\m/\m_0)$ as
\begin{equation}
\label{RGinvB}
B_\p(\m/\m_0) = B_\p^{\rm RG}E^+(\m/\m_0)\ .
\end{equation}
The new LEC, $B_\p^{\rm RG}$, is both RG invariant and scale invariant,
also by assumption.
Hence $B_\p(\m/\m_0) m(\m/\m_0)=B_\p^{\rm RG}m_0$, and
using Eq.~(\ref{barea}) it follows that under a scale transformation
\begin{equation}
\label{dBmdl}
  \frac{\partial}{\partial \log\l} B_\p(\m/(\l\m_0))\, m(\l;\m)
  = + B_\p(\m/(\l\m_0))\, m(\l;\m) \ .
\end{equation}
The factor $E^-(e^{\t}f_\p/\m_0)$ in Eq.~(\ref{calBpion})
is invariant under a scale transformation
by construction, because the combination $e^{\t}f_\p/\m_0$ is.\footnote{%
  Being $\m$ independent, $E^-(e^{\t}f_\p/\m_0)$ is trivially RG invariant.
}
Noting that the scaling dimension of $\S$ is zero,
and taking the contribution from the factor $e^{3\t}$ into account,
we obtain
\begin{equation}
\label{dAllmdl}
  \frac{\partial}{\partial \log\l}\,\cl_m\Big|_{\l=1}
  = 4 \cl_m +x_\m\frac{\partial}{\partial x_\m}\cl_m
  =\frac{\partial}{\partial x_\m}\left(x_\m\cl_m\right)\ ,
\end{equation}
which establishes the invariance of the action.
This conclusion  is valid for any choice of the function $\g_m$.

The lagrangian for dChPT with a varying $\g_m$ function is given
by Eq.~(\ref{LOlag}), with now $\cl_m$ given by Eq.~(\ref{calBpion}).
The theory is invariant under the scale transformation
of the effective fields, Eq.~(\ref{diltransf}), combined with the
spurion transformation rules\footnote{%
  In Ref.~\cite{PP} we introduced a space-time dependent spurion field $\c(x)$
  for the renormalized mass, but for our present purposes,
  a space-time independent spurion for $m$ is sufficient.
}
\begin{subequations}
\label{transeff}
\begin{eqnarray}
\label{transeffa}
m(\m) &\to& m(\l;\m)\ ,\\
\label{transeffb}
\m_0 &\to& \l\m_0\ ,\\
\label{transeffc}
c_0 &\to& c_0-\log\l \ ,\\
\label{transeffd}
c_1 &\to& c_1 \ .
\end{eqnarray}
\end{subequations}
The transformation rule~(\ref{transeffc}) is needed to ensure the invariance
of (the space-time integral of) $\cl_d$ in Eq.~(\ref{potentialsa}).\footnote{%
  The transformation rules of $c_0$ and $c_1$ get modified at higher orders.
  For a detailed discussion of $\cl_d$, see Refs.~\cite{PP,largemass}.}
As usual,
once the spurions $m$, $\m_0$ and $c_0$ are set equal to their fixed values,
this breaks the scale symmetry explicitly.

We may again shift the $\t$ field,
as we did in Sec.~\ref{tree},  such that after the shift it has a
vanishing expectation value for $m=0$.   The LECs $f_{\p,\t}$ and
$B_\t$ are redefined as in Eq.~(\ref{shiftlec}), but now $\hB_\p$ is defined as
\begin{equation}
\label{Bhatpi}
\hB_\p(\m/\m_0) = \hB_\p^{\rm RG} E^+(\m/\m_0)\ , \qquad
\hB_\p^{\rm RG} = e^{v_0} B_\p^{\rm RG}\ ,
\end{equation}
so that $\hB_\p = e^{v_0}B_\p$.
The lagrangian after the shift is again given by Eq.~(\ref{LOlagredef}),
but now with
\begin{equation}
\label{calBpionredef}
\cl_m=-\half \hf_\p^2\, e^{3\t}\, E^-(e^{\t}\hf_\p/\m_0)\,
\hB_\p(\m/\m_0) m(\m/\m_0)\, \tr(\S+\S^\dagger) \ ,
\end{equation}
instead of Eq.~(\ref{hatpotb}).  Note that, instead of being a function
of $e^{\t} f_\p/\m_0$, now $E^-$ is a function of $e^{\t}\hf_\p/\m_0$.

Let us now reconsider the trace anomaly.
We first apply the scale transformation only to the effective fields,
setting the spurions equal to their fixed values.  In this case,\footnote{%
  We omit the contribution from the scale dependence of
  the space-time coordinates (compare Eq.~(\ref{dAllmdl})).}
\begin{equation}
\label{taul}
\frac{\partial}{\partial\log\l}
= \frac{\partial\t}{\partial\log\l}\,\frac{\partial}{\partial \t}
=\frac{\partial}{\partial \t} \ ,
\end{equation}
and we obtain the contribution of $\cl_m$ to $\partial_\m S_\m$,
the divergence of the dilatation current $S_\m$ (see App.~D of Ref.~\cite{PP}),
\begin{equation}
\label{TmmE}
  \left(\frac{\partial}{\partial\t} - 4\right) \cl_m
  = -(1+\g_m(e^{\t}\hf_\p/\m_0)) \cl_m
  = -(1+\g_m(e^{\t}\hf_\p/\m_0)) m \bj\j({\rm EFT}) \ .
\end{equation}
In the last step we identified $\cl_m$ with the EFT representation of
$m \bj\j$ in the underlying theory.   This reproduces, in the EFT, the
contributions from the fermions to the trace anomaly \cite{CDJ}.
Recall that we have defined $\g_m$ to be a function of $\m/\m_0$,
\seef Eq.~(\ref{gammag}).  Replacing $\t$ by $v(m)$, its vacuum expectation value
at non-vanishing $m$, we see that Eq.~(\ref{TmmE}) effectively identifies
the renormalization scale $\m$ with $F_\p=e^{v(m)}\hf_\p$, \seef Eq.~(\ref{quantb}).
This reveals a key feature of our construction of $\g$-dChPT:
$\g_m$ is evaluated at a renormalization scale equal to the physical scale
$F_\p$, which, in turn, is a function of the input fermion mass.
We comment that we chose the hadronic scale
inside $E^-$ in Eq.~(\ref{calBpion}) to be $f_\p$, but,
to achieve the desired scaling behavior, we could equivalently
choose $f_\t$, or, more generally, any other hadronic scale $m_h$
that enters the dChPT lagrangian (or generalization thereof)
via the combination $e^{\t} m_h$, such as, for example, the nucleon mass
in the chiral limit.

We now specialize to specific choices for the function $\g_m$.
First, for constant $\g_m=\g_*$,
\begin{equation}
\label{contact}
\hB_\p(\m/\m_0) E^-(e^\t \hf_\p/\m_0)
=  \hB_\p^{\rm RG}
\left(\frac{\m}{\m_0}\right)^{\g_*} e^{-\g_*\t}\left(\frac{\m_0}{\hf_\p}\right)^{\g_*}
=  \hB_\p(\m/\hf_\p)\, e^{-\g_*\t}\ ,
\end{equation}
and the lagrangian $\cl_m$ in Eq.~(\ref{calBpionredef})
reduces to Eq.~(\ref{hatpotb}).\footnote{%
  In this special case, the dependence on $\m_0$ drops out.}
This also implies $\hB_\p(\m/\hf_\p) = e^{(1-\g_*)v_0} B_\p(\m/f_\p)$,
consistent with Eq.~(\ref{shiftlecc}).

We next introduce a new choice for $\g_m$ that we will be using
for the actual fits to the KMI data.
With $t=\t+\log(\hf_\p/\m_0)$ we define
\begin{equation}
\label{EFtilde}
E^-(e^\t \hf_\p/\m_0) = E^-(e^t)=e^{-\tilde{F}(t)}\ ,
\end{equation}
where
\begin{equation}
\label{Ftilde}
\tilde{F}(t) = \tilde\g_0 t - \half\,\tilde{b} t^2
+ \frac{1}{3}\,\tilde{c} t^3\ ,
\end{equation}
a cubic polynomial in $t$.
The variable $t$ is invariant under scale transformations, and,
consistent with our general discussion, $\tilde\g_0$, $\tilde{b}$
and $\tilde{c}$ are LECs that do not depend on $\m$ or $\m_0$.
Re-expressing $t$ in terms of $\t$, we write
\begin{eqnarray}
\label{FtF}
\tilde{F}(t) &=& \tilde{F}(\log(\hf_\p/\m_0)) + F(\t) \ ,
\\
\label{F}
F(\t) &=& \g_0 \t - \half\,b \t^2 + \frac{1}{3}\,c \t^3\ ,
\end{eqnarray}
which defines the coefficients of the cubic polynomial $F(\t)$
in terms of those of $\tilde{F}(t)$, and $\log(\hf_\p/\m_0)$.
Substituting into Eq.~(\ref{calBpionredef}),
and absorbing $e^{-\tilde{F}(\log(\hf_\p/\m_0))}$ into $\hB_\p$,
the final form of the lagrangian becomes
\begin{equation}
\label{calBpionFfinal}
\cl_m = -\half \hf_\p^2  \hB_\p m\,e^{3\t-F(\t)}  \,\tr(\S+\S^\dagger) \ .
\end{equation}

We will use the acronym $\g$-dChPT for the lagrangian defined by Eq.~(\ref{LOlag}),
with $\cl_d$ given by Eq.~(\ref{potentialsa}), and $\cl_m$ by Eq.~(\ref{calBpionFfinal})
for some general function $F(\t)$.  Of course, for the case of a linear $F(\t)$,
Eq.~(\ref{calBpionFfinal}) reduces to Eq.~(\ref{potentialsb}),
and the lagrangian is just LO dChPT.

As an EFT, dChPT is based on the power counting established in
Refs.~\cite{PP,largemass} and reviewed above.
As in ordinary ChPT, loop corrections in dChPT can be included systematically;
the power counting~(\ref{pc}) dictates which terms occur at the next-to-leading
order (NLO) \cite{PP}, at the next-to next-to-leading order (NNLO),
and so on.  The same is true in the large-mass regime,
where the power counting is controlled by Eq.~(\ref{LMexpand}).
This raises the question of how much $\g$-dChPT deviates
from the strict EFT framework of dChPT itself.
If we rely on algebraic structure and symmetries only,
this allows $E^-(e^{\t}\hf_\p/\m_0)$ in Eq.~(\ref{calBpionredef}),
or, equivalently, $F(\t)$ in Eq.~(\ref{calBpionFfinal}), to depend on
an infinite number of parameters, reflecting the model nature of $\g$-dChPT.
But if, on the other hand, we assume that $F(\t)$ takes the form of Eq.~(\ref{F}),
with
\begin{equation}
\label{bcsmall}
\g_0 \sim (n_f-n_f^*)^0=1\ ,\qquad b\sim n_f-n_f^*\ ,\qquad c\sim (n_f-n_f^*)^2\ ,
\end{equation}
then the factor $e^{-F(\t)}$ may be obtained via partial resummation
of terms from all orders in the expansion in powers of $n_f-n_f^*$.  It thus
reflects a fairly modest departure from dChPT,
in that we will be taking into account some higher-order analytic terms,
resummed into $e^{-F(\t)}$, while omitting other higher-order terms.
In addition, we will not calculate any non-analytic higher-order corrections
when fitting $\g$-dChPT to data.  We will re-examine the scenario
of Eq.~(\ref{bcsmall}) after presenting our fits to the KMI data
in Sec.~\ref{KMI}.

%%%%%%%%%%%%%%%%
\vspace{3ex}
\begin{boldmath}
\subsection{\label{hadrons} Hadronic quantities for varying $\g_m$}
\end{boldmath}
%%%%%%%%%%%%%%%%
As in Sec.~\ref{tree}, we begin with the saddle-point equation.
For $m\ge 0$ the potential is minimized
by setting $\S=1$ in Eq.~(\ref{calBpionredef}), and $v=v(m)$ is the solution of
(compare Eq.~(\ref{saddle}))
\begin{equation}
\label{spgeneral}
\frac{(3-\g_m)m}{4c_1\cm}\, = v e^{v+F(v)}\ ,
\end{equation}
where now
\begin{equation}
\label{gammaF}
  \g_m = F'(v) \ .
\end{equation}
When $F(\t)$ is linear in $\t$ we reproduce the results of Sec.~\ref{tree},
whereas for $F(\t)$ in Eq.~(\ref{F}) we have
\begin{equation}
\label{gammacub}
\g_m = \g_0 -b v + c v^2\ .
\end{equation}

Equation~(\ref{spgeneral}) can be rewritten as
\begin{equation}
\label{sprew}
m = \frac{d_2}{\td_1}\,\frac{1}{3-\g_m}\,v\,e^{v+F(v)}\ ,
\end{equation}
with
\begin{equation}
\label{dtilde}
\td_1 = \frac{\hf_\p^2}{8\hB_\p c_1\cm}\ .
\end{equation}
For a general function $F$,
Eq.~(\ref{sprew}) cannot be explicitly inverted analytically.
We will, in effect, solve it numerically for $m$ as a function of $v$,
as described in Sec.~\ref{KMI}.
In terms of $v$, $F_\p$ is still given by Eq.~(\ref{quantb}).
The pion mass is now
\begin{equation}
\label{Mpi}
M_\p^2 = 2 \hB_\p m \,e^{v-F(v)}\ ,
\end{equation}
so that, using Eq.~(\ref{sprew}),
the ratio $M_\p^2/F_\p^2$ is given by
\begin{equation}
\label{MFratio}
\frac{M_\p^2}{F_\p^2}=
\frac{1}{\td_1}\,\frac{v}{3-\g_m}\ .
\end{equation}
The three equations~(\ref{sprew}),~(\ref{quantb}) and~(\ref{MFratio}) contain six
parameters, $\td_1$, $d_2$, $\hf_\p$ and
the three parameters inside $F$: $\g_0$, $b$ and $c$.

We will not fit $M_\t$ to the KMI data, as the errors found in Ref.~\cite{LatKMI}
are too large for such a fit to have statistical relevance.
We will, however, fit the staggered taste-splittings obtained
in Ref.~\cite{LatKMI}.
With $M_{\G_i}$ the masses of the taste-split pions corresponding to the tastes
\begin{equation}
\label{tastelist}
\G_i\in\{\G_5,\G_{\m 5},\G_{\m\n},\G_\m,\G_I\}\ ,
\end{equation}
we will fit the differences\footnote{
  We note that $M_{\G_5}=M_\p$ is the mass of the Nambu--Goldstone pion.}
\begin{equation}
\label{tastesplittings}
\D(\G_i) \equiv a^2(M_{\G_i}^2 - M_\p^2)\ ,
\end{equation}
according to  \cite{LS,AB}
\begin{subequations}
\label{tsplit}
\begin{eqnarray}
\label{tsplitP}
  \D(\G_5) &\equiv& \D_P \ = \ 0 \ ,
\\
\label{tsplitA}
  \D(\G_{\m 5}) &\equiv& \D_A \ = \
   C_1 E(\g_1) + 3 C_3 E(\g_3) + \phantom{1} C_4 E(\g_4) + 3 C_6 E(\g_6) \ ,
  \hspace{8ex}
\\
\label{tsplitT}
  \D(\G_{\m\n}) &\equiv& \D_T \ = \
  \hspace{11ex} 2 C_3 E(\g_3) + 2 C_4 E(\g_4) + 4 C_6 E(\g_6) \ ,
\\
\label{tsplitV}
  \D(\G_\m) &\equiv& \D_V \ = \
  C_1 E(\g_1) + \phantom{1} C_3 E(\g_3) + 3 C_4 E(\g_4) + 3 C_6 E(\g_6) \ ,
\\
\label{tsplitS}
  \D(\G_I) &\equiv& \D_S \ = \
  \hspace{11.2ex} 4 C_3 E(\g_3) + 4 C_4 E(\g_4) \ .
\end{eqnarray}
\end{subequations}
Here $C_{1,3,4,6}$ are LECs associated with the taste-breaking potential \cite{AB},
and
\begin{equation}
\label{Edef}
E(\g_i) = e^{(4-\g_i)v} \ .
\end{equation}
Equation~(\ref{Edef}) assumes that $\g_i$, the anomalous dimensions of the
taste-breaking four-fermion operators, are constant
(see Ref.~\cite{GNS} for more details).
A global fit of the data including all the taste splittings
has eight new parameters, coming from Eq.~(\ref{tsplit}),
in addition to the six parameters of the basic fit.
This is a large number of parameters, and, as we will see,
some of them are not sufficiently constrained by the available data.
Thus, we will not venture into an exploration of any scale dependence
of the $\g_i$.

We end this section with a comment.
While in LO dChPT the potential is bounded from below,
in $\g$-dChPT with general $F(v)$
the potential can be unbounded from below.\footnote{
  For polynomial $F(v)$, a necessary and sufficient condition that
  the potential will be bounded from below is that the highest power of $v$
  is even, and its coefficient is positive.
}
Mathematically, this appears to be a problem,
but we contend that it is physically irrelevant.   Within the EFT
framework, the potential can only be known for $\co(1)$ values of the fields.
While the pion field is always $\co(1)$ because it is a compact field, this is
not the case for $\t$.   We thus need to restrict the EFT to $\co(1)$ values of
$\t$ ``by hand.''  In practice, this means that after fits to the data,
we need to check that indeed values of $v$ predicted by the fits are $\co(1)$,
and do not land in the large-field region.
In all our fits with a varying $\g_m$ indeed
unphysical regions of the potential occur at very large values of $v$,
but they are separated from the physical region by an
exponentially large potential barrier.  Consistently, our fits never explore
the unphysical region of the potential.

%%%%%%%%%%%%%%%%%%%%%%%%%%%
\vspace{3ex}
%\newpage
\section{\label{KMI} Fits to the L\lowercase{at}KMI data}
%%%%%%%%%%%%%%%%%%%%%%%%%%%
In this section, we will present our fits to data reported in Ref.~\cite{LatKMI},
obtained by the LatKMI collaboration for the eight-flavor SU(3) gauge theory.
We begin in Sec.~\ref{data} with a discussion of these data
and the policies we will follow when we use them.
In Sec.~\ref{window}, we present ``window'' fits.
These are fits of $M_\p^2/F_\p^2$ and $aF_\p$ to the predictions
of LO dChPT, for successive quintets of fermion masses,
from the five lightest masses to the five heaviest ones.
Altogether, ten different fermion masses were simulated in Ref.~\cite{LatKMI},
making six (overlapping) windows.
The window fits test the constancy of the LO dChPT parameters.
We find a systematic trend of change for all fit parameters, by much more
than their errors allow, proving that the full KMI mass range cannot be fit
to LO dChPT.
Then, in Sec.~\ref{varying} we fit the data at all ten fermion masses
simultaneously to $\g$-dChPT, the extension of LO dChPT with a varying $\g_m$
constructed in Sec.~\ref{vary}, with the special choice of $\g_m$ in
Eq.~(\ref{gammacub}). We find that this extension of dChPT
successfully describes the KMI data set.
Data for taste-split pion masses is available
for a more limited set of fermion masses, and we present
our fits including the taste splittings in Sec.~\ref{taste}.
We end with a discussion of the scale dependence of $\g_m$
found in our fits in Sec.~\ref{KMIdisc}.

The simulations of Ref.~\cite{LatKMI} were all performed at the same bare coupling.
Invoking a mass-independent scale setting prescription,
this implies that all ensembles have a common lattice spacing $a$.
We elaborate on the choice of a scale setting prescription
in App.~\ref{massindep}.

We will be using lattice units in all our fits.
This means taking $\m=\m_0=1/a$, and thus $m(\m)=m(\m_0)=m_0$.

%%%%%%%%%%%%%%%%%%%%%%%%%%%
%\newpage
\subsection{\label{data} The LatKMI data}
%%%%%%%%%%%%%%%%%%%%%%%%%%%
The pion mass $M_\p$ and decay constant $F_\p$ were measured
in Ref.~\cite{LatKMI} at ten bare-mass values
\begin{equation}
\label{barem}
am_0\in\{0.012,\ 0.015,\ 0.02,\ 0.03,\ 0.04,\ 0.05,\ 0.06,\ 0.07,\ 0.08,\ 0.1\}\ .
\end{equation}
In Ref.~\cite{LatKMI} a great effort was made to also determine
the dilaton mass $M_\t$.  It was found that indeed a dilaton exists,
roughly degenerate with the pions.
$M_\t$ was measured for only 6 fermion masses,
leaving out $am_0=0.05$, $0.07$, $0.08$ and $0.1$.
More seriously, the statistical errors of $M_\t$ turn out to be
too large to have any real impact on our fits.
In the window fits to LO dChPT (next subsection),
we found that when we include a fit of $M_\t^2/F_\p^2$ to Eq.~(\ref{quantd})
in our global fit, $d_3$ remains largely undetermined,
while all other fit parameters
do not change.  The only noticeable change is a higher $p$-value,
as might be expected.
We thus omit the dilaton mass from the fits discussed in this paper.

Other hadron masses were also determined,
notably the vector meson mass $aM_\r$ and the nucleon mass $aM_N$.\footnote{
  The pions are too heavy for the $\r$ to decay.
}
For these hadrons, the prediction from LO dChPT
is that the ratios $M_\r/F_\p$ and $M_N/F_\p$ should be
independent of $am_0$ \cite{largemass};
this is also true if we extend LO dChPT to include a varying $\g_m$.
Excluding the two largest fermion masses, $am_0=0.08$ and 0.1,
we found that we can fit $M_\r/F_\p$ to a constant, with a $p$-value of 0.31.
$M_N$ was measured only for a subset of the fermion masses,
\begin{equation}
\label{barem2}
am_0\in\{0.012,\ 0.015,\ 0.02,\ 0.03,\ 0.04,\ 0.06,\ 0.08\}\ ,
\end{equation}
which leaves out $am_0=0.05$, $0.07$ and 0.1.
Keeping only the 5 lightest masses,
we found that a fit of $M_N/F_\p$ to a constant has a $p$-value of 0.07.
This suggests that for larger fermion masses, higher-orders corrections
in dChPT (other than a varying $\g_m$) would be needed to fit these ratios.
In addition, discretization effects could be playing a bigger role
(see below).  We will thus focus in this paper
on the pion sector, considering
$M_\p^2/F_\p^2$ and $aF_\p$ in Secs.~\ref{window} and \ref{varying},
and adding taste splittings in Sec.~\ref{taste}.

Information on the systematic errors of $aM_\p$ and $aF_\p$ is incomplete.
Mostly, they were measured on at least two different volumes,
and we estimate the finite-volume error by taking the difference between the
results at the largest two volumes.
For $am_0=0.012$ only one volume is available.  In this case
we took the finite-volume errors to be the same as for $am_0=0.015$.
The latter was simulated on the same volume as $am_0=0.012$,
as well as on a somewhat smaller volume.
We note that, since $am_0=0.012$ is the lightest fermion mass,
this procedure may underestimate its finite-volume errors.
A single volume was reported also for $am_0=0.08$ and 0.1.
For these fermion masses, the two largest ones, $M_\p L$ is very large,
and finite-volume corrections should be very small.  We thus took
the finite-volume errors for these two masses to vanish.
We added the statistical error and the finite-volume error
of $aM_\p$ and $aF_\p$ in quadrature.
These errors were propagated to the ratio $M_\p^2/F_\p^2$,
and correlations between this ratio and $aF_\p$ were kept.\footnote{%
  Correlations between $aM_\p$ and $aF_\p$ on each ensemble
  are not available.  We note that, in Ref.~\cite{GNS}, we found that
  these correlations are small in the LSD data.}

As the simulations of Ref.~\cite{LatKMI} were done at a single bare coupling,
no direct information is available on the lattice spacing dependence,
and it is not possible to take the continuum limit.
We are thus forced to ignore scaling violations in our fits,
but it should be kept in mind that these affect our results
in an unknown way.
Generally speaking, $M_\r$ and $M_N$ are larger than $M_\p$,
and are thus prone to larger discretization effects.
Also, as an example, for $am_0=0.08$ Ref.~\cite{LatKMI} finds
the central values $aM_\p=0.51$, $aM_\r=0.68$ and $aM_N=1.02$,
hence, at the largest fermion masses discretization effects
could be significant for the pions as well.
We will briefly mention evidence for scaling violations
in the determination of the gradient flow scale $t_0$ in Sec.~\ref{KMIdisc}.
The only other information on lattice spacing effects comes from
pion taste splittings.   The masses of taste-split pions, which were measured
only on the seven ensembles with bare masses~(\ref{barem2}),
will be considered in Sec.~\ref{taste}.

%%%%%%%%%%%%%%%%%%%%%%%%%%%
%\newpage
\subsection{\label{window} Window fits}
%%%%%%%%%%%%%%%%%%%%%%%%%%%
We begin with fitting $M_\p^2/F_\p^2$ and $aF_\p$ to the predictions
of LO dChPT, Eqs.~(\ref{quanta}) and~(\ref{quantb}).  We consider
sets of five successive
fermion masses, taking first the lightest five masses from the set~(\ref{barem}),
then the second to the sixth masses, \etc, for a total of six quintets.
The results are shown in Table~\ref{KMIwindow}.\footnote{%
  We will label fits with a number for the table, and a letter for the
  fit in the table.   For example, fit 1A refers to fit A in Table 1, \etc
}
All the fits are good.   However, the parameter values change
with the partial mass range, more than allowed by their errors.
In particular, the lowest mass range (fit \ref{KMIwindow}A)
and the highest mass range  (fit \ref{KMIwindow}F) do not overlap,
hence their parameter errors are statistically independent.
These fits are thus not consistent with each other.
A simultaneous fit of LO dChPT to all ten masses
has a $p$-value of order $10^{-11}$.
Clearly, the whole KMI mass range cannot be fit to LO dChPT.

As dChPT admits a systematic expansion, the failure to describe
a set of data at LO means that higher orders in the expansion are needed.
However, already at LO, dChPT contains more parameters than ordinary ChPT.
Depending on the observables being fitted, many more would be needed
for an NLO fit.  We believe that much better data
is required for a meaningful NLO fit.
As discussed in Sec.~\ref{data}, the LSD and KMI data sets both contain
only a single lattice spacing, leaving discretization errors as an
uncontrolled source of systematic uncertainty.
In addition, it may well be that more refined data,
for additional bare masses and/or with smaller statistical errors,
would be needed to determine all the parameters in the NLO fit.

%%%%%%%%%%%%%%%%%%%%%%%%%%%%%%%%%%%%%%%%%%%%%%%%%%%%%%%%%%%%%%%%%%%%%%%%%%%%%%%
\begin{table}[t]
\begin{center}
%\hspace{-0.3cm}
\begin{ruledtabular}
\begin{tabular}{|c|c|c|c|c|c|c|}
%\hline
& A & B & C & D & E & F \\
\hline
range & 0.012--0.04 & 0.015--0.05 &0.02--0.06 & 0.03--0.07 & 0.04--0.08 & 0.05--0.1 \\
\hline
$\c^2$/dof & 9.37/6  & 9.85/6 & 4.81/6 & 4.38/6 & 4.56/6 & 3.83/6 \\
$p$-value  & 0.15 & 0.13 & 0.57 & 0.63 & 0.60 & 0.70  \\
\hline
$\g_*$  & 0.608(8) & 0.589(10) & 0.543(10) & 0.534(12) & 0.527(8) & 0.498(13) \\
$a\hf_\p$   &0.0050(7)& 0.0067(6)& 0.0089(8) & 0.010(2) & 0.011(1) & 0.011(1) \\
$\td_1$ & 0.0716(44) & 0.0629(28) & 0.0545(23) & 0.0512(56)&0.0500(28) &0.0484(28)  \\
$-\log(ad_2)$  & 10.5(3) & 10.0(2) & 9.5(2) & 9.2(4) & 9.1(2) & 9.0(2)  \\
%\hline
\end{tabular}
\end{ruledtabular}
\end{center}
%\vspace*{4ex}
\floatcaption{KMIwindow}{{\it Fits of the KMI data to Eqs.~(\ref{quanta})
and~(\ref{quantb}), using selections of five successive fermion masses
from the set~(\ref{barem}).
All parameter errors reported in this paper are hessian.}}
\end{table}
%%%%%%%%%%%%%%%%%%%%%%%%%%%%%%%%%%%%%%%%%%%%%%%%%%%%%%%%%%%%%%%%%%%%%%%%%%%%%%%

%%%%%%%%%%%%%%%%%%%%%%%%%%%
%\newpage
\begin{boldmath}
\subsection{\label{varying} Fits with a varying $\g_m$}
\end{boldmath}
%%%%%%%%%%%%%%%%%%%%%%%%%%%
Being unable to carry out a full NLO fit at present, we are left with
the option of partially extending LO dChPT by exploring different
``directions'' in ``higher-order parameter space.''  By its very nature,
no such extension is fully systematic, and each extension should thus
be considered a model.  Our assumption is that our model, $\g$-dChPT, captures
the relevant physics better than other extensions of LO dChPT.

As we have discussed in Sec.~\ref{hypers}, the physical mechanism that
underlies the behavior of the LSD data is hyperscaling.
The KMI mass range is higher than the LSD one, which motivates us to consider
a minimal modification of this physical picture.  We assume that
the KMI mass range is still governed by the same principles that produce
hyperscaling in the LSD mass range, except that, because of the diminishing
influence of the IRFP, we now have to allow the mass anomalous dimension
to vary.  That consideration has led us to the framework of $\g$-dChPT,
developed in Sec.~\ref{vary}.

In this subsection, we will thus consider fits of the KMI data to $\g$-dChPT.
Specifically, we consider fits of $M_\p^2/F_\p^2$ and $aF_\p$ to Eqs.~(\ref{MFratio})
and~(\ref{quantb}), where $\g_m$ is quadratic in $v$, \seef Eq.~(\ref{gammacub}).
We begin with a technical issue.
The independent variable in these equations is $v$, which, in turn,
can be determined in terms of $am_0$ using Eq.~(\ref{sprew}).
However, unlike in LO dChPT discussed in Sec.~\ref{tree},
Eq.~(\ref{sprew}) cannot be analytically inverted.\footnote{%
  In principle, the formal inverse function $m=m(v)$ may not be
  single valued.  In practice, we found that $v$ is monotonically increasing
  with $m$ over the entire KMI mass range.}
Instead, in addition to the parameters defining the $\g$-dChPT lagrangian,
we introduce new parameters $v_i$, one per ensemble.\footnote{%
  The total number of parameters increases by the number of $v_i$ parameters,
  \ie, by the number of ensembles included in the fit.
  The number of data increases by the same amount (the $am_{0,i}$),
  leaving the number of degrees of freedom unchanged.
}
We fit the corresponding bare mass $am_{0,i}$ to Eq.~(\ref{sprew}),
while simultaneously also fitting $(M_\p^2/F_\p^2)_i$ and $(aF_\p)_i$,
all as functions of the same parameter $v_i$.
Artificially introducing a tiny error for $am_{0,i}$,
the fit in effect solves Eq.~(\ref{sprew}) numerically
for $v_i$ in terms of $am_{0,i}$.
Thus, for given values of the $\g$-dChPT parameters,
$v_i$ is equal to $v(am_{0,i})$ with numerical precision
set by the ``error'' of the ``data'' $am_{0,i}$.
We have varied the errors on $am_{0,i}$ between $10^{-6}$ and $10^{-7}$,
finding no discernible differences in the results of our fits.
$\c^2$ values remain equal to four decimal places,
whether one includes the ``$am_0$ part'' in the computation of $\c^2$ or not.

As in Ref.~\cite{GNS}, we can calculate $(a\hB_\p)_i$ on each ensemble
using Eq.~(\ref{Mpi}) and our fit result for $v_i$.
In all cases studied in this paper the so-obtained values of $(a\hB_\p)_i$
are equal within error.  This confirms the self-consistency
of our assumption that the lattice spacing $a$ is independent
of the fermion mass.

%%%%%%%%%%%%%%%%%%%%%%%%%%%%%%%%%%%%%%%%%%%%%%%%%%%%%%%%%%%%%%%%%%%%%%%%%%%%%%%
\begin{table}[t]
\begin{center}
\begin{ruledtabular}
\begin{tabular}{|c|c|c|c||c|}
%\hline
& A & B & C & D \\
\hline
omitted & --- & 0.1 & 0.1, 0.08 & 0.1, 0.08\\
\hline
$\c^2$/dof &  20.7/14 & 11.5/12 & 10.0/10 & 14.8/11 \\
$p$-value  & 0.11 & 0.48  & 0.44& 0.19 \\
\hline
$\hf_\p$   & 0.0104(4) & 0.0102(5) & 0.0101(9) & 0.0085(5) \\
$\td_1$ &  0.0506(10) & 0.0512(12) & 0.0516(23) & 0.0559(17)\\
$-\log(ad_2)$  & 10.1(2) & 10.4(2) & 10.6(3) & 9.9(1)\\
$\g_0$  & 1.69(23) & 2.11(31)  & 2.29(57) & 0.85(5) \\
$b$ & 0.97(22) & 1.38(31) & 1.57(67) & 0.12(2) \\
$c$ & 0.20(5) & 0.30(8) & 0.35(16) & ---\\
%\hline
\end{tabular}
\end{ruledtabular}
\end{center}
%\vspace*{4ex}
\floatcaption{LatKMIgamma}{\it Fits of $M_\p^2/F_\p^2$ and $aF_\p$ to
$\g$-dChPT, the extension of LO dChPT discussed in Sec.~\ref{vary}.
The ``omitted'' row shows bare mass values from the set~(\ref{barem})
which are not included in the fit, if any.
}
\end{table}
%%%%%%%%%%%%%%%%%%%%%%%%%%%%%%%%%%%%%%%%%%%%%%%%%%%%%%%%%%%%%%%%%%%%%%%%%%%%%%%

The results of our fits are shown in Table~\ref{LatKMIgamma}.
Fit \ref{LatKMIgamma}A includes all ten ensembles,
fit \ref{LatKMIgamma}B leaves out the $am_0=0.1$ ensemble,
and fit \ref{LatKMIgamma}C leaves out both $am_0=0.1$ and 0.08.
All the fits are good, but fits \ref{LatKMIgamma}B and \ref{LatKMIgamma}C
are better than fit \ref{LatKMIgamma}A.
We also carried out fits setting $c=0$, \ie, taking $\g_m$ in Eq.~(\ref{gammacub})
to be a linear function of $v$.  Fits with $c=0$ including all ten ensembles,
or omitting the $am_0=0.1$ ensemble, have very low $p$-values,
0.001 and 0.01 respectively.  We do not show them in the table.
However, if we omit both the $am_0=0.1$ and $0.08$ ensembles,
we obtain fit \ref{LatKMIgamma}D, which is a good fit.
The parameters $a\hf_\p$, $\td_1$ and $\log(ad_2)$ are relatively stable
between the fits with $c$ as a free parameter,
and fit \ref{LatKMIgamma}D, where $c=0$.
By contrast, the parameters defining the function $\g_m$ change substantially:
Fit \ref{LatKMIgamma}D yields much smaller values for both $\g_0$ and $b$
than the other fits of Table~\ref{LatKMIgamma}.

The results of fits \ref{LatKMIgamma}B and \ref{LatKMIgamma}D
are shown in Fig.~\ref{varyfit}.  The black points are data
that were included in the fits, whereas the magenta points were excluded.
The lower left panel shows that if we simplify our {\it ansatz} for $\g_m$
to be linear in $v$, then the $am_0=0.08$ and 0.1 ensembles
must be excluded.

%%%%%%%%%%%%%%%%%%%%%%%%%%%%%%%%%%%%%%%%%%%%%%%%%%%%%%%%%%%%%%%%%%%%%%%%%%%%%%%%
\begin{figure}[t!]
\vspace*{4ex}
\begin{center}
\includegraphics*[width=7cm]{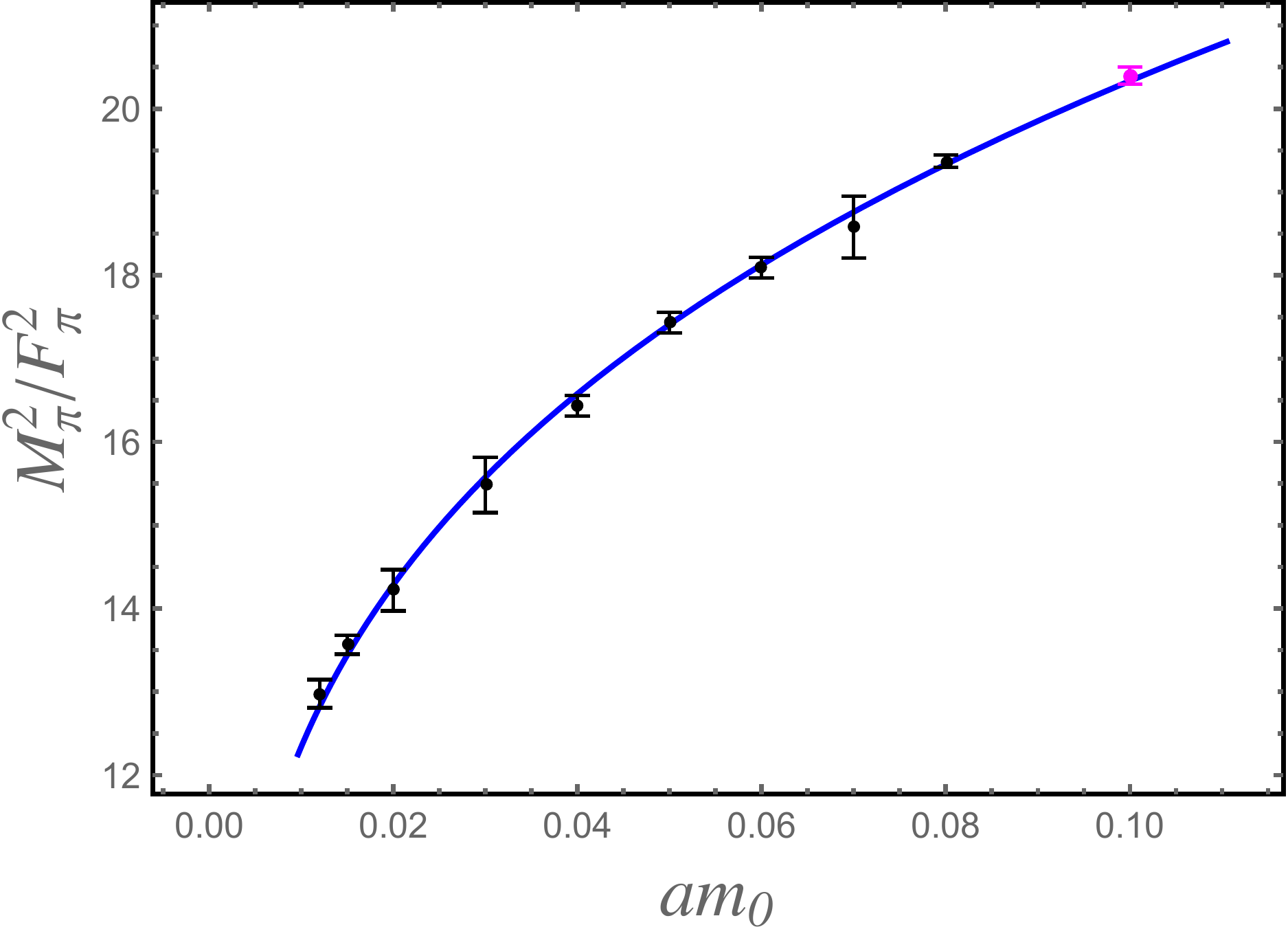}
\hspace{0.5cm}
\includegraphics*[width=7cm]{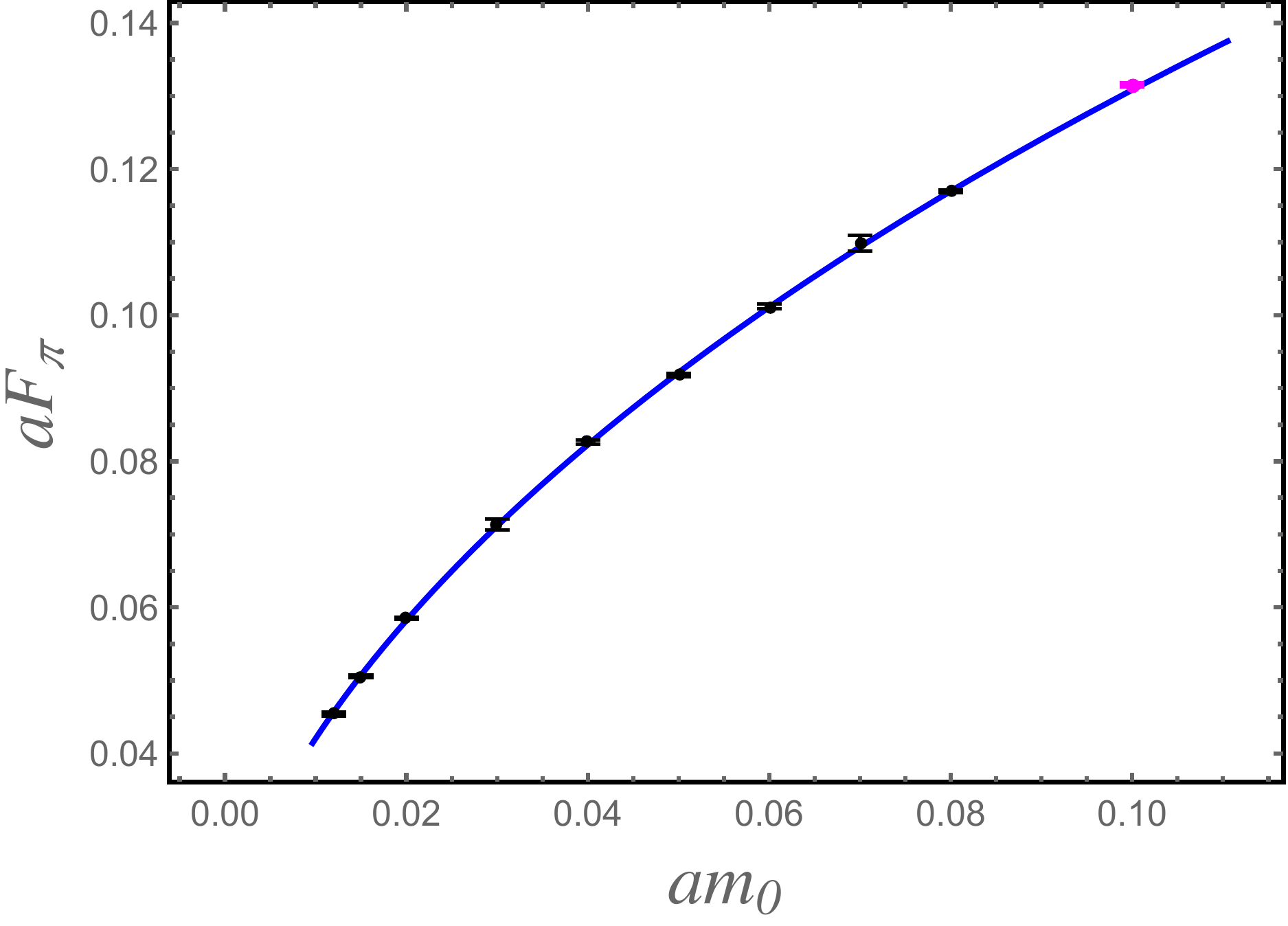}
\includegraphics*[width=7cm]{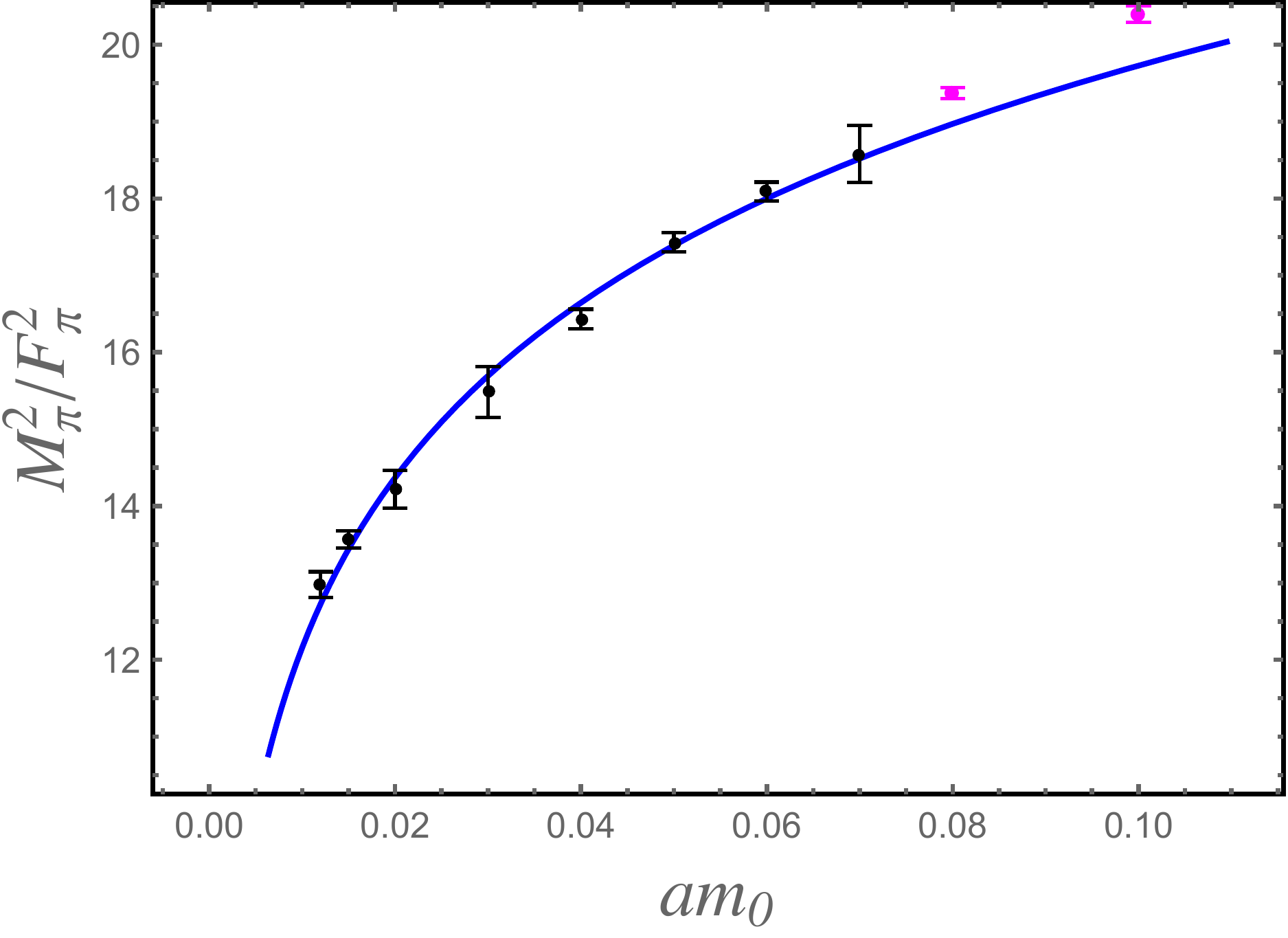}
\hspace{0.5cm}
\includegraphics*[width=7cm]{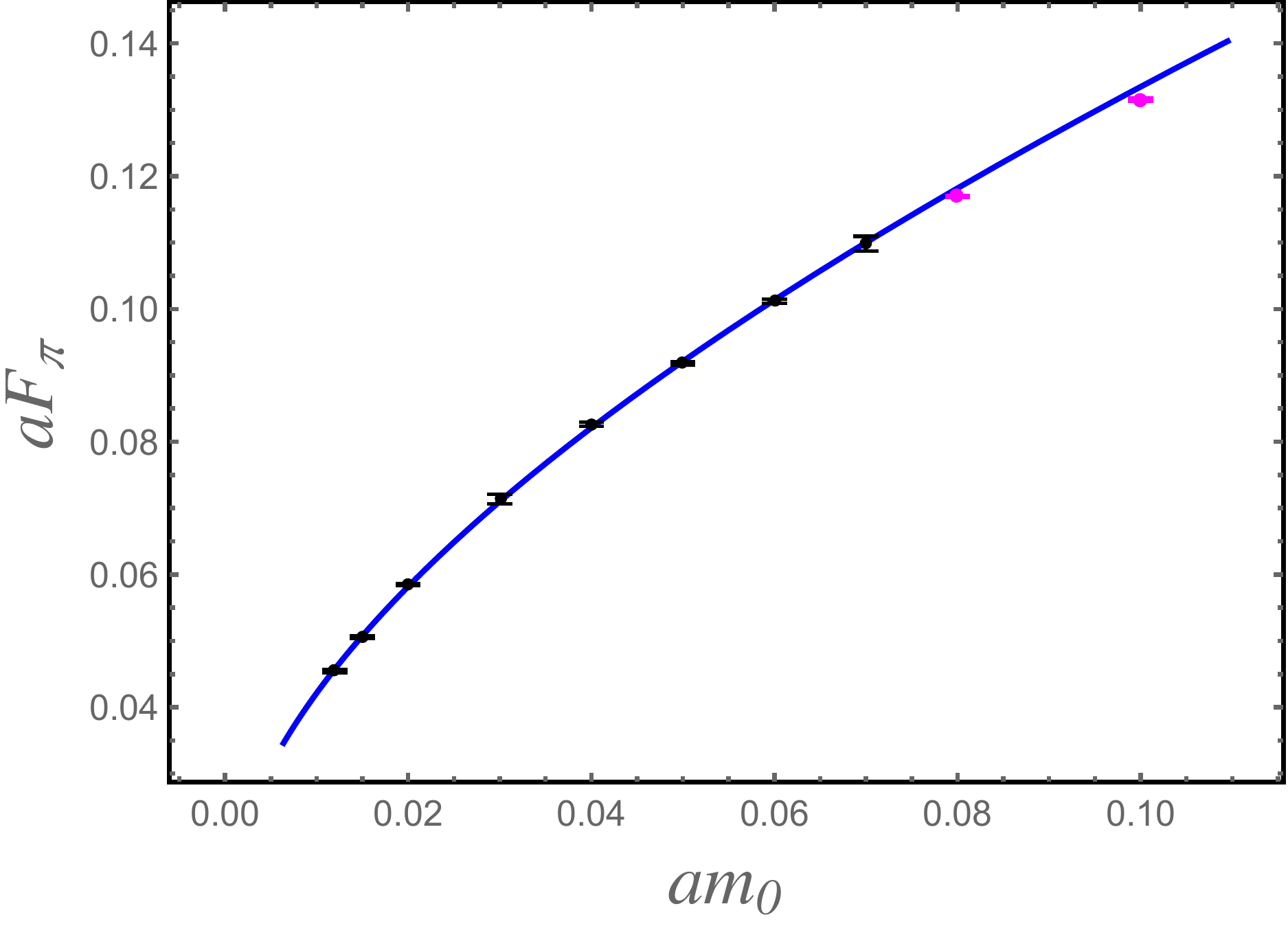}
\end{center}
\begin{quotation}
\floatcaption{varyfit}%
{\it Upper panels: Fit results for $M_\p^2/F_\p^2$ (left panel)
and $aF_\p$ (right panel) using fit~\ref{LatKMIgamma}B.
Lower panels: similar, using fit~\ref{LatKMIgamma}D.
Black points are fitted data, while magenta points
were not included in the fit.}
\end{quotation}
\vspace*{-4ex}
\end{figure}
%%%%%%%%%%%%%%%%%%%%%%%%%%%%%%%%%%%%%%%%%%%%%%%%%%%%%%%%%%%%%%%%%%%%%%%%%%%%%%%%

We have proposed in Sec.~\ref{vary} that the exponential factor
$e^{-F(v)}$ may originate from a resummation of the dominant contributions from
all orders in the expansion in $n_f-n_f^*$.  According to
the hypothesis~(\ref{bcsmall}),
$b$ is an NLO parameter, while $c$ is an NNLO parameter.
One way to test this scenario is to examine the effect of truncating
the Taylor expansion of the exponential factor.
The range of values we find for $v$ in the fits to the KMI data
is $1.5\le v\le 2.5$.  Considering first fit \ref{LatKMIgamma}B,
we can compare the numerical values of
$\mbox{exp}\left(\half b v^2-\frac{1}{3}c v^3\right)$, and
its version truncated at NNLO, namely
$1+\half b v^2+\frac{1}{8}b^2v^4-\frac{1}{3}cv^3$.
When we vary $v$ from 1.5 to 2.5, the exponential and its truncated  version
take values ranging from 3.4 to 15, respectively 3.4 to 13.
The differences (taking the correlations into account) are $-0.05(8)$
and 2(4), respectively, so that the exponential and truncated forms
are consistent with each other.
The situation is somewhat different for fit \ref{LatKMIgamma}D,
where the smallness of both $b$ and its relative error
allows for a more precise comparison.
Varying again $v$ from 1.5 to 2.5, $\mbox{exp}\left(\half b v^2\right)$ varies
from 1.15 to 1.48, while the expansion to NLO, $1+\half b v^2$,
varies from 1.14 to 1.39. The (correlated) differences are 0.010(4) and 0.09(3),
respectively.  Thus, while the behavior of both forms is qualitatively similar,
the differences are statistically significant.
Fits with the truncated version give results consistent with
fits \ref{LatKMIgamma}B and \ref{LatKMIgamma}D, but with
lower $p$-values.

Without more data it is difficult to decide which fit
in Table~\ref{LatKMIgamma} is the preferred one.
Clearly, unless the two heaviest masses are dropped,
$c$ must be kept in the fit.  Given its (conjectural) role as an
NNLO parameter, it is to be expected that eventually $c$ will be needed
to describe the data as the mass range is increased.
Still, we cannot rule out that the main reason why
fit \ref{LatKMIgamma}D does not accommodate the two heaviest masses
is large scaling violations at those mass values.

In all fits where the parameter $c$ is present, it is always small
compared to $b$, consistent with the conjectured hierarchy~(\ref{bcsmall}).
However, in the same fits, one cannot say that $b$ is small compared to $\g_0$.
By contrast, in fit \ref{LatKMIgamma}D, where $c=0$,
also $b$ is clearly small compare to $\g_0$.  The most appealing scenario
thus appears to be the following.  We exclude the two largest fermion
mass values, because they require going to (at least) NNLO
in the EFT expansion,
and/or because they are afflicted by too large scaling violations.
The remaining mass range may be amenable
to an NLO dChPT fit,\footnote{%
  With the caveats discussed in Sec.~\ref{window}.
}
for which fit \ref{LatKMIgamma}D is our closest substitute.

%%%%%%%%%%%%%%%%%%%%%%%%%%%%%%%%%%%%%%%%%%%%%%%%%%%%%%%%%%%%%%%%%%%%%%%%%%%%%%%
\begin{table}[t]
\begin{center}
\begin{ruledtabular}
\begin{tabular}{|c|c|c||c|c|c|}
%\hline
& A & B & C & D & E \\
\hline
omitted & --- & --- & --- & --- & 0.08 \\
\hline
$\c^2$/dof &  17.5/28 & 38.2/34 &  29.5/29 & 50.1/35 & 22.9/29\\
$p$-value  & 0.94 & 0.28 & 0.44 & 0.05 & 0.78 \\
\hline
$\hf_\p$   & 0.0102(5) &0.0105(4) & 0.0095(4) & 0.0099(4) & 0.0085(6) \\
$\td_1$ &  0.0512(12) & 0.0503(10) & 0.0526(12) & 0.0514(10) & 0.0562(20) \\
$-\log(ad_2)$  & 10.3(2) & 10.2(2) & 9.68(8) & 9.65(7) & 9.98(13) \\
$\g_0$  & 1.85(27) & 1.81(26) & 0.82(3) & 0.86(2) & 0.92(3) \\
$b$ & 1.12(28) &1.09(27) &  0.121(17) & 0.142(10) & 0.154(17) \\
$c$ & 0.24(7) & 0.23(7) & --- & ---& ---\\
\hline
$-\log{C_1}$ & 10(10) & --- & $-$6(45) & --- & --- \\
$\g_1$ & 4(6) & --- & 14(29) & --- & --- \\
$-\log{C_3}$ & 11(4) & --- &  8(8) & --- & ---\\
$\g_3$ & 3(2) & --- & 5(5) & --- & ---\\
$-\log{C_4}$ & 12.3(3) & 12.1(2) & 12.6(3) & 12.3(2) & 12.5(3) \\
$\g_4$ & 1.29(10) & 1.36(8) & 1.26(10) & 1.34(8) & 1.42(11)  \\
$-\log{C_6}$ & 18(6) & --- & 18(4) & --- & --- \\
$\g_6$ & 0(2) & --- & 0(2) & --- & --- \\
%\hline
\end{tabular}
\end{ruledtabular}
\end{center}
%\vspace*{4ex}
\floatcaption{tastes}{\it Fits of $M_\p^2/F_\p^2$, $aF_\p$ and taste splittings
to $\g$-dChPT.  The ``omitted'' row shows bare mass values
from the set~(\ref{barem2}) which are not included in the fit, if any.
For description see text.
}
\end{table}
%%%%%%%%%%%%%%%%%%%%%%%%%%%%%%%%%%%%%%%%%%%%%%%%%%%%%%%%%%%%%%%%%%%%%%%%%%%%%%%

%%%%%%%%%%%%%%%%%%%%%%%%%%%
%\newpage
\vspace{3ex}
\subsection{\label{taste} Taste splittings}
%%%%%%%%%%%%%%%%%%%%%%%%%%%
We now turn to fits which also include the
taste splittings~(\ref{tastesplittings}), \ie, fits
of $M_\p^2/F_\p^2$, $aF_\p$ and $\D_{A,T,V,S}$ to $\g$-dChPT,
augmented by Eq.~(\ref{tsplit}).  Our fits are limited to the smaller
ensemble set~(\ref{barem2}), where the taste-split pion masses
were measured.

We show five different fits in Table~\ref{tastes}.
Fit~\ref{tastes}A includes all the parameters:
the basic $\g$-dChPT parameters of Sec.~\ref{varying}, namely
$a\hf_\p$, $\td_1$, $\log(ad_2)$, $\g_0$, $b$ and $c$,
as well as all eight taste-splitting parameters of Eq.~(\ref{tsplit}).
Data from all seven ensembles in the set~(\ref{barem2})
are included in the fit.   The $p$-value is very high.
The results for the six basic $\g$-dChPT parameters
are consistent with fit \ref{LatKMIgamma}B.\footnote{%
Note that the ensemble set~(\ref{barem2}) does not include $am_0=0.1$.}
As for the taste-splitting parameters, most of them, namely,
$\g_{1,3,6}$ and $\log{C_{1,3,6}}$, are not well determined by the fit.
We conclude that fit~\ref{tastes}A gives an
excellent description of the data, but the data are not precise enough
to determine all parameters in the fit.

We next consider fits omitting poorly determined parameters.
Among the taste-splitting parameters,
only $\log{C_4}$ and $\g_4$ were determined with good precision.
As for $C_1$, $C_3$ and $C_6$,
if we take their errors seriously, using them as 1$\s$ bounds,
these parameters are ``allowed'' to be very small relative to $C_4$
(by factors $\sim 2\times 10^{3}$, $\sim 10$ and $\sim 10^{5}$, respectively).
Setting $C_1=C_3=C_6=0$, we obtain fit~\ref{tastes}B.  This is a good fit,
even though its $p$-value is much smaller than fit~\ref{tastes}A,
as one would expect.  The results of fits~\ref{tastes}A and~\ref{tastes}B
are in very good agreement.
The dominance of the taste splittings generated by the $C_4 E(\g_4)$ term
is consistent with the results we obtained for the LSD data \cite{GNS},
as well as with the familiar taste splittings found in QCD.

In Sec.~\ref{varying} we saw that the parameter $c$ can be omitted if
the fermion masses $am_0=0.1$ and 0.08 are not included in the fit.
While $am_0=0.08$ is present in the ensemble set~(\ref{barem2}),
we also repeated fits~\ref{tastes}A and~\ref{tastes}B while setting $c=0$,
obtaining fits~\ref{tastes}C and~\ref{tastes}D, respectively.
Finally, fit~\ref{tastes}E is similar to fit~\ref{tastes}D,
except that the $am_0=0.08$ ensemble is not included.  Fit~\ref{tastes}C,
were we set $c=0$ but keep all the taste-splitting parameters,
is very good.  Setting both $c=0$ and $C_1=C_3=C_6=0$ leads to a relatively low
$p$-value in fit~\ref{tastes}D.  After dropping the $am_0=0.08$ ensemble,
in fit~\ref{tastes}E the $p$-value is again very high.

Our results for $a\hf_\p$, $\td_1$, $\log(ad_2)$ are fairly consistent in
all the fits reported in Tables~\ref{LatKMIgamma} and~\ref{tastes}.
The values of the parameters defining the function $\g_m$
are consistent among the fits where $c\ne 0$: fits~\ref{LatKMIgamma}A,
\ref{LatKMIgamma}B, \ref{LatKMIgamma}C, \ref{tastes}A and~\ref{tastes}B.
In the fits with $c=0$ the values of $\g_0$ and $b$ are different,
but again consistent across this group: fits~\ref{LatKMIgamma}D,
\ref{tastes}C, \ref{tastes}D, and~\ref{tastes}E.
The values of the taste splitting parameters $\log C_4$ and $\g_4$
are consistent in all the fits of Table~\ref{tastes}, while the
(poorly determined) values of the remaining taste splitting parameters
are consistent between fits~\ref{tastes}A and~\ref{tastes}C.

%%%%%%%%%%%%%%%%%%%%%%%%%%%%%%%%%%%%%%%%%%%%%%%%%%%%%%%%%%%%%%%%%%%%%%%%%%%%%%%%
\begin{figure}[t]
\vspace*{4ex}
\begin{center}
\includegraphics*[width=10cm]{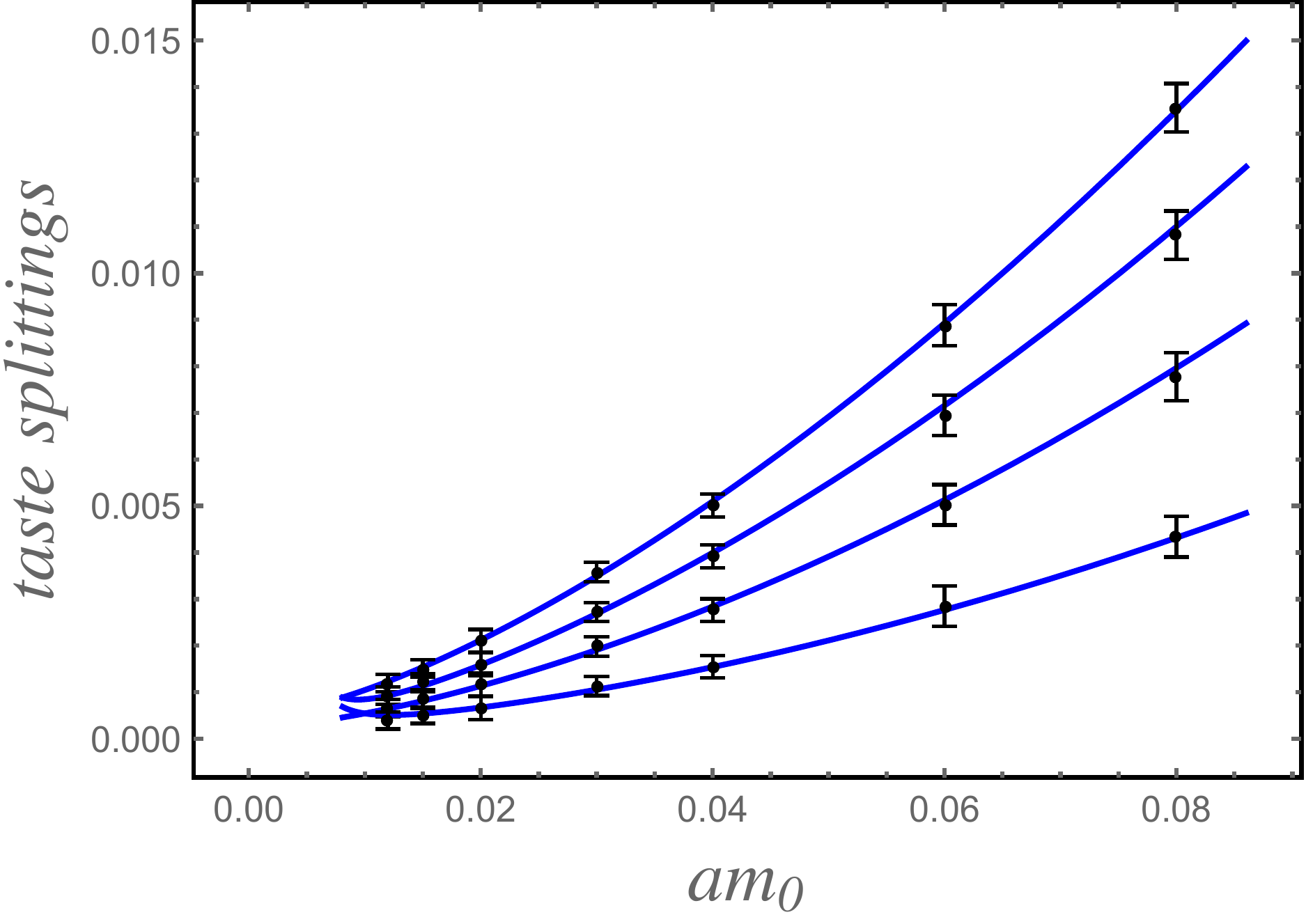}
\end{center}
\begin{quotation}
\floatcaption{tasteplot}%
{{\it Fit \ref{tastes}C of the taste splittings $\D_{A,T,V,S}$ of Eq.~(\ref{tsplit}),
as a function of $am_0$.
From top to bottom: $\D_S$, $\D_V$, $\D_T$, and $\D_A$.}}
\end{quotation}
\vspace*{-4ex}
\end{figure}
%%%%%%%%%%%%%%%%%%%%%%%%%%%%%%%%%%%%%%%%%%%%%%%%%%%%%%%%%%%%%%%%%%%%%%%%%%%%%%%%

In fit \ref{tastes}C, LO dChPT has been minimally extended
(within the framework of $\g$-dChPT)
to include an NLO correction to the function $\g_m$.  This fit
gives an excellent description of the ensemble set~(\ref{barem2})
with taste splittings included;
the parameter $c$ is not needed.   We thus consider
fit \ref{tastes}C to be the preferred fit from Table~\ref{tastes}.
We plot the taste splittings of this fit in Fig.~\ref{tasteplot}.
A caveat is that, even though all the taste-split pion masses
were measured in Ref.~\cite{LatKMI}, the data are not precise enough
to determine all taste-splitting parameters.\footnote{
  By contrast, the LSD data, which we fitted in Ref.~\cite{GNS},
  contains only $M_{\m 5}$ and $M_{\m\n}$ \cite{LSD2}.}
We recall that the QCD taste splittings are essentially independent
of the fermion mass \cite{MILC,AB}.\footnote{%
  Thanks to the dominance of $C_4$, the QCD taste splittings
  are also roughly equal to each other.}
By contrast, as for the LSD data \cite{GNS},
also in the KMI mass range the taste splittings vary with the fermion mass.
This behavior can be successfully described in dChPT,
where the scale dependence of the taste-breaking operators gives rise
to mass dependent tree-level taste splittings,
through the factors $E(\g_i)$ in Eq.~(\ref{tsplit}).

%%%%%%%%%%%%%%%%%%%%%%%%%%%
%\newpage
\begin{boldmath}
\subsection{\label{KMIdisc} Scale dependence of $\g_m$}
\end{boldmath}
%%%%%%%%%%%%%%%%%%%%%%%%%%%
The anomalous dimension function $\g_m$ obtained from two of the fits
of Table~\ref{LatKMIgamma} is shown in Fig.~\ref{gammaplot}.
The blue band represents fit~\ref{LatKMIgamma}B, where $\g_m = F'(v)$
is quadratic in $v$ (Eq.~(\ref{gammacub})), while the magenta band
represents fit~\ref{LatKMIgamma}D, where $\g_m$ is linear in $v$.
With Eq.~(\ref{quantb}), we take the argument of $\g_m$
to be $v=\log(aF_\p/a\hf_\p)$, and then plot $\g_m$ as a function of $aF_\p$.
The two $\g_m$ functions agree well in most of the interval containing
the fitted data, $0.045\;\ltap\;aF_\p\;\ltap\;0.12$.
The good agreement deteriorates towards the lower end of the interval,
below which these functions diverge from each other.
If we would overlay the (constant) results of each window fit
from Sec.~\ref{window} as a set of horizontal bands (each stretching
over its corresponding range of $aF_\p$), these bands would
be consistent with the blue and magenta bands in that interval.\footnote{%
  We do not show window fits in Fig.~\ref{gammaplot} because
  the different bands become visually difficult to see.
}

%%%%%%%%%%%%%%%%%%%%%%%%%%%%%%%%%%%%%%%%%%%%%%%%%%%%%%%%%%%%%%%%%%%%%%%%%%%%%%%
\begin{figure}[t!]
\vspace*{4ex}
\begin{center}
\includegraphics*[width=10cm]{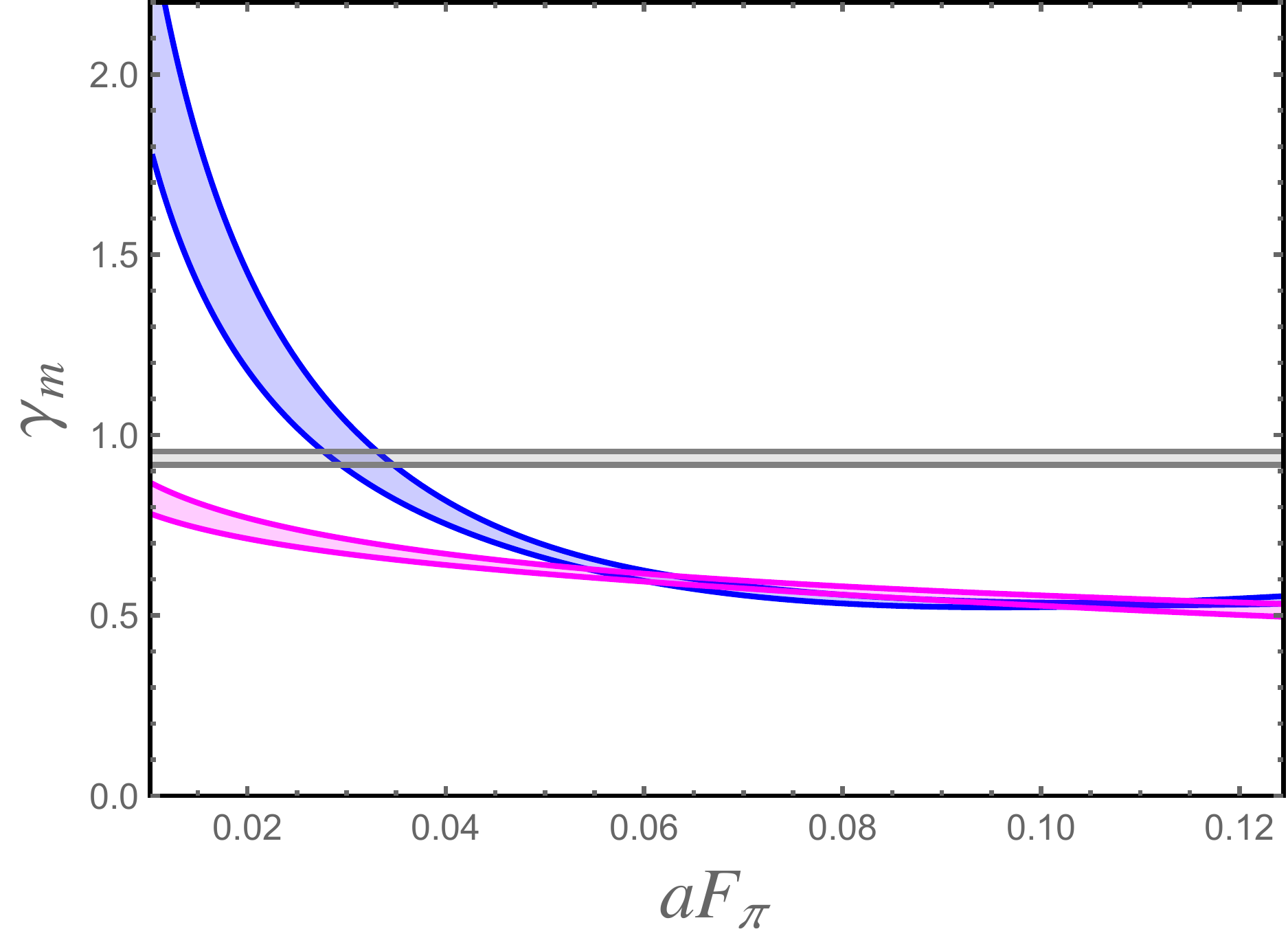}
\end{center}
\begin{quotation}
\floatcaption{gammaplot}%
{{\it The running mass anomalous dimension $\g_m$,
obtained from fit~\ref{LatKMIgamma}B (blue band) and
\ref{LatKMIgamma}D (magenta band), plotted as a function of $aF_\p$
(see text).  The gray horizontal band is $\g_*=0.936\pm 0.019$,
from our fit to the LSD data \cite{GNS}.
The fitted KMI data have values of $aF_\p$ between $0.045$ and $0.12$.}}
\end{quotation}
\vspace*{-4ex}
\end{figure}
%%%%%%%%%%%%%%%%%%%%%%%%%%%%%%%%%%%%%%%%%%%%%%%%%%%%%%%%%%%%%%%%%%%%%%%%%%%%%%%

Figure~\ref{gammaplot} also shows the value $\g_*=0.936(19)$ obtained from
our fits of the LSD data to LO dChPT \cite{GNS}, as a gray horizontal band.
The LSD mass range is lower than the KMI range, and
the (generalized) hyperscaling behavior we have observed implies that
the LSD range of $F_\p$ should also be lower than the corresponding
KMI range, in physical units.  Equivalently, the LSD values of $aF_\p$,
properly converted to KMI lattice units, should lie to the left
of the KMI range of $aF_\p$ in Fig.~\ref{gammaplot}.

Since the LSD data is successfully described by a constant $\g_m=\g_*$,
we expect that also in the chiral limit $\g_m$ will remain constant,
at a value consistent with $\g_*$.
The continuity of $\g_m$ as a function of $F_\p$ thus requires that,
as $F_\p$ is lowered from the KMI range into the LSD range,
$\g_m$ will rise to a value consistent with $\g_*$,
and then stay roughly constant all the way to the chiral limit.
It is intriguing that the strong dynamics of the $N_f=8$ system
might induce this behavior of $\g_m$.\footnote{%
  A $\g_m$ function that saturates to a constant value at strong coupling
  was observed in the SU(2) theory with two adjoint Dirac fermions \cite{SU2}.
}
Fig.~\ref{gammaplot} shows that, when extrapolated below the KMI range,
the quadratic $\g_m$ of fit~\ref{LatKMIgamma}B overshoots $\g_*$,
while the linear $\g_m$ of fit~\ref{LatKMIgamma}D undershoots it.
The desired behavior of $\g_m$ over the combined KMI and LSD ranges
cannot be described by simple {\em ansatzes} such as the ones we have used.
One cannot rule out, however, that the combined LSD and KMI mass ranges could be
described by including higher orders in dChPT systematically.

Clearly, an investigation of the combined LSD and KMI mass ranges would be
extremely interesting.  However, this is just not possible with
the existing data sets.  We already pointed out that the LSD and KMI
data sets were each produced at a single lattice spacing.
Moreover, the lattice actions used by LSD and by KMI differ in their details,
and scaling violations can potentially differ significantly between
the two lattice actions and axial currents.
This means that the only way to reliably compare these results
is by first taking the continuum limit separately for the LSD lattice action
and for the KMI lattice action.  The minimal requirement to make this possible
is a second set of data at a different lattice spacing,
for each lattice action.\footnote{%
  To make sure that the same physical mass range is covered,
  one can, for example, monitor the values of some observable,
  such as a hadron mass or a decay constant, in units of $\sqrt{t_0}$.
}

We have attempted a comparison of the LSD and KMI lattice scales,
using $t_{0,{\rm ch}}$, the chiral-limit value of the gradient-flow scale $t_0$
\cite{MLflow}, which we have determined for the LSD data set in Ref.~\cite{GNS}.
The comparison is deficient for several reasons.
First, unlike in ordinary ChPT \cite{BG},
dChPT does not predict the behavior of $t_0$ as a function of
the fermion mass \cite{GNS}, so the best we can do is a phenomenological fit.
Second, usually the gradient flow scale (or its chiral limit)
is used to compare the lattice spacings of ensembles generated with
different bare couplings, but with the same lattice action.
By contrast, here we are comparing results obtained using two different
lattice actions, hence the meaning of the comparison is less clear.
Finally, there are also scaling violations in the lattice observables used
to extract $t_0$, as well as in the gradient-flow equation.
KMI used two lattice definitions for $t_0$
which should agree in the continuum limit, but which
consistently differ by some 15\% over the entire KMI mass range;
we do not have equivalent information about uncertainties associated
with the LSD data.  With all these caveats in mind,
our findings suggest that the ratio $r=a({\rm KMI})/a({\rm LSD})$
is smaller than one.  Using Eq.~(\ref{barem}) together with Eq.~(\ref{LSDmasses}) below,
it follows that the KMI mass range is indeed higher than the LSD mass range,
in agreement with the physical picture reflected in Fig.~5 of Ref.~\cite{LSD}.
But, we are unable to turn this conclusion into a more quantitative statement.

We close this section with a comment.  As discussed above,
our experimentation with $t_0$ (and its chiral extrapolation) suggests that
$r<1$.  Now, an alternative way to estimate $r$ would be to take advantage
of the fact that $\hf_\p$, the chiral-limit value of the pion decay constant,
is a physical observable.
Expecting $\sqrt{2}\hf_\p({\rm LSD}) \approx \hf_\p({\rm KMI})$
in physical units,\footnote{%
  The factor of $\sqrt{2}$ is due to different normalization conventions.
}
it follows that $a\hf_\p({\rm KMI})/(\sqrt{2}a\hf_\p({\rm LSD})) \approx r$.
The reason why we only expect an approximate equality between
$\sqrt{2}\hf_\p({\rm LSD})$ and $\hf_\p({\rm KMI})$,
is the different scaling violations of the two lattice actions.
In reality, using the value of $a\hf_\p({\rm LSD})$ from Ref.~\cite{LSD},
and taking $a\hf_\p({\rm KMI}) \sim 0.01$, we find
$a\hf_\p({\rm KMI})/(\sqrt{2}a\hf_\p({\rm LSD})) \sim 10$, in stark conflict
with the estimate $r<1$ obtained from the gradient flow scale.  It is unlikely
that scaling violations {\em per-se} can account for this inconsistency.
The problem must be related to the long extrapolation
to the chiral limit inherent in the extraction of $a\hf_\p$.
It does not necessarily imply that ($\g$-)dChPT cannot be trusted.
The factor $e^{v(m)} = F_\p/\hf_\p$ is very sensitive to $m$,
which makes a long extrapolation to the chiral limit
much more difficult than in the case of QCD.
For at least one of the data sets our fit result
for $a\hf_\p$ is likely to contain a large, and unaccounted for,
source of systematic error.  A comparison of the values of $ad_2$
obtained from the two data sets reveals a similar,
and, in fact, more severe, problem,
which presumably have a similar source, given that $d_2 = \hf_\p^2/(2\hB_\p)$.
We comment that in order to compare $a\hB_\p$ between the LSD and KMI
lattice scales we have to apply an RG transformation, but once again,
it is hard to see how such a transformation would suffice to match
the values of $d_2$ found in the two simulations.

%%%%%%%%%%%%%%%%%%%%%%%%%%%
%\newpage
\vspace{3ex}
\begin{boldmath}
\section{\label{potential} The $\D$ class of dilaton potentials}
\end{boldmath}
%%%%%%%%%%%%%%%%%%%%%%%%%%%
So far, we have considered a model modification of the
LO dChPT form of $\cl_m$, based on the observation that
the coupling of the underlying theory may start running
at the physical scale determined by a growing fermion mass, thereby
inducing a varying mass anomalous dimension as well.
In this section we turn to a class of modifications to
the dilaton-potential term $\cl_d$.
Alternate forms of the dilaton potential were first applied to
the LSD data in Ref.~\cite{AIP1}.   In Ref.~\cite{AIP3}
a class of dilaton potentials $\cl_\D$ was proposed, defined by
(compare Eq.~(\ref{hatpot}))
\begin{subequations}
\label{VDelta}
\begin{eqnarray}
  \cl_\D(\t) &=& \hf_\t^2 \hB_\t e^{4\t}\, V_\D(\t) \ ,
\label{VDeltaa}\\
  V_\D(\t) &=& \frac{c_1}{4-\D} \left(1-\frac{4}{\D}\, e^{(\D-4)\t} \right) \ ,
\label{VDeltab}
\end{eqnarray}
\end{subequations}
where $\D$ is a new free parameter.\footnote{%
  $\cl_\D(\t)$ is bounded from below for any $-\infty < \D < \infty$.
}
We have translated the notation of Ref.~\cite{AIP3} to our notation.
In the limit $\D\to 4$, the potential $\cl_d$ of Eq.~(\ref{hatpota}) is recovered.
For $\D=2$, $\cl_\D$ becomes the linear $\s$-model potential
considered in Ref.~\cite{Kutietal}.  We will refer to the low-energy lagrangian
with $\cl_d$ replaced by $\cl_\D$ as $\D$-dChPT.

Applying $\D$-dChPT to the LSD data,
Ref.~\cite{AIP3} concluded that these data appear to favor a value of $\D$
around $3.5$, with a large uncertainty.  Correlations in these data were not
taken into account \cite{AIP3}.
Moreover, correlations which occur because of
the appearance of $F_\p$ in all three equations fitted in Ref.~\cite{AIP3},
as well as the appearance of $M_\p$ in two of them,
apparently were not taken into account either.
In Sec.~\ref{Delta} we begin by collecting the expressions needed to fit
$\D$-dChPT.  In Sec.~\ref{LSD} we revisit the determination of $\D$
using the LSD data, taking all correlations into account.
This analysis departs from the framework of LO dChPT (Sec.~\ref{tree})
only by replacing the dilaton potential $\cl_d$  by $\cl_\D$.
At this stage
the mass anomalous dimension is held fixed, \seef\ Eq.~(\ref{potentialsb}).
Then, in Sec.~\ref{KMIDelta}, we explore fits of the KMI data to the $\D$ class
of potentials.  As in the previous section, we consider both fixed-$\g_m$ fits
to subsets of the KMI data, as well as fits with a varying $\g_m$ to
the entire KMI data set.  We summarize our findings in Sec.~\ref{disc}.

Unlike the modification of $\cl_m$ to accommodate a running $\g_m$,
we are not aware of a concrete physical motivation to replace $\cl_d$
by the more general form $\cl_\D$.  A closely related question is
whether or not $\D$-dChPT is the leading order in a systematic
low-energy expansion for an arbitrary value of $\D$.

The potential $\cl_d$, Eq.~(\ref{hatpota}), is based on the
systematic power counting developed in Ref.~\cite{PP}.
Since $\cl_d$ corresponds to the limit $\D\to 4$ in Eq.~(\ref{VDelta}),
it follows by continuity that there must exist a neighborhood of $\D=4$ where
the dChPT systematic expansion is still applicable.
For arbitrary $\D$, a power counting was proposed in Ref.~\cite{AIP3}.
We prove in App.~\ref{power counting} that the arguments given in Ref.~\cite{AIP3}
are not correct.  $\D$-dChPT, \ie, the
low-energy lagrangian consisting of Eq.~(\ref{tree}) with $\cl_d$ replaced by
$\cl_\D$, should thus be considered to be a model.

%%%%%%%%%%%%%%%%%%%%%%%%%%%
%\newpage
\vspace{3ex}
\begin{boldmath}
\subsection{\label{Delta} Fitting data to $\cl_\D$}
\end{boldmath}
%%%%%%%%%%%%%%%%%%%%%%%%%%%
For the case of a constant $\g_m=\g_*$,
combining Eq.~(\ref{VDelta}) with $\cl_m$ of Eq.~(\ref{potentialsb}),
one finds the saddle-point equation relating $v$ to $m$,
\begin{equation}
\label{DsaddleM}
m =\frac{d_2}{d_1} \frac{1-e^{(\D-4)v}}{4-\D}\, e^{(1+\g_*)v} \ .
\end{equation}
It is then straightforward to derive the relations
\begin{subequations}
\label{quantDelta}
\begin{eqnarray}
\frac{M_\p^2}{F_\p^2}&=&\frac{1}{d_1} \frac{1 - e^{(\D-4)v}}{4-\D}
\equiv h_\D(m) \ ,
\label{quantDeltaa}\\
F_\p&=& \hf_\p e^{v}
\label{quantDeltab}\\
    &=&\left(\frac{d_0 m}{h_\D(m)}\right)^{\frac{1}{1+\g_*}} \ ,
\label{quantDeltac}\\
\frac{M_\t^2}{F_\p^2} &=&d_3 \Big(1 + (\D+\g_*-3) d_1 h_\D(m) \Big) \ ,
\label{quantDeltad}
\end{eqnarray}
\end{subequations}
where we used the definitions~(\ref{ds}).

In the case of a varying $\g_m$, Eq.~(\ref{quantDeltab}) is still applicable,
while combining Eq.~(\ref{VDelta}) with $\cl_m$ of Eq.~(\ref{calBpionFfinal}),
Eqs.~(\ref{DsaddleM}) and~(\ref{quantDeltaa}) generalize to
\begin{subequations}
\label{generalize}
\begin{eqnarray}
\label{generalizea}
m&=&\frac{d_2}{\td_1}\, \frac{1-e^{(\D-4)v}}{4-\D}\, \frac{e^{v+F(v)}}{3-\g_m}\ ,
\\
\label{generalizeb}
\frac{M_\p^2}{F_\p^2}&=&\frac{1}{\td_1(3-\g_m)}\,\frac{1-e^{(\D-4)v}}{4-\D}\ ,
\end{eqnarray}
\end{subequations}
where $\g_m$ is given in Eq.~(\ref{gammaF}),
and $\td_1$ is defined in Eq.~(\ref{dtilde}).

We now turn to fits of the LSD and KMI data, in order to explore to what extent
they constrain the value of $\D$.   We emphasize again that this investigation
is empirical, as no systematic power counting is available for this model
for arbitrary values of $\D$.

%%%%%%%%%%%%%%%%%%%%%%%%%%%%%%%%%%%%%%%%%%%%%%%%%%%%%%%%%%%%%%%%%%%%%%
\begin{table}[t]
\begin{center}
\begin{ruledtabular}
\begin{tabular}{|c|c|c||c|c|}
%\hline
& A & B & C & D \\
\hline
omitted & --- & 0.00889 & --- & 0.00889 \\
\hline
$\c^2$/dof & 8.72/9  & 2.50/6  & 15.18/13  & 5.52/8 \\
$p$-value  & 0.56    & 0.87    & 0.30      & 0.70   \\
\hline
$\D$        & 2.8(7)   & 3.5(7)  & 2.7(6)    & 3.5(7)    \\
$\g_*$      & 0.935(19)  & 0.936(19)   & 0.933(19)   & 0.937(19)   \\
$\log{d_0}$ & 1.94(6)  & 1.93(6)   & 1.94(6)   & 1.93(6)   \\
$d_1$       & 0.042(20)  & 0.083(84)   & 0.037(15)   & 0.073(66)   \\
$-\log(ad_2)$ & 11.6(9) & 12.9(2.5) & 11.3(7)  & 12.6(2.1) \\
$d_3$       & 17(9) & 9(9)   & 20(8)  & 10(9)   \\
$-\log{C_1}$ & ---             & ---              & ---       & ---    \\
$\g_1$       & ---             & ---              & ---       & ---    \\
$-\log{C_3}$ & ---             & ---              & 9.7(6)    & 10(2)  \\
$\g_3$       & ---             & ---              & 2.0(1)    & 2.4(7) \\
$-\log{C_4}$ & ---             & ---              & 8.3(7)    & 10(2)  \\
$\g_4$       & ---             & ---              & 1.96(6)   & 2.1(4) \\
$-\log{C_6}$ & ---             & ---              & 36(7)     & 17(11) \\
$\g_6$       & ---             & ---              & $-$11(4)  & 0(3)   \\
%\hline
\end{tabular}
\end{ruledtabular}
\end{center}
%\vspace*{4ex}
\floatcaption{tabD}{{\it Fits of the LSD data to $\D$-dChPT.
The fits to the right of the double vertical line include taste breaking;
those to the left do not.  The ``omitted'' row shows bare mass values
from the set~(\ref{LSDmasses}) which are not included in the fit, if any.
}}
\end{table}
%%%%%%%%%%%%%%%%%%%%%%%%%%%%%%%%%%%%%%%%%%%%%%%%%%%%%%%%%%%%%%%%%%%%%%

%%%%%%%%%%%%%%%%%%%%%%%%%%%%
\vspace{3ex}
\subsection{\label{LSD} The LSD data}
%%%%%%%%%%%%%%%%%%%%%%%%%%%%
Data reported in Ref.~\cite{LSD2} includes results at five different fermion masses,
\begin{equation}
\label{LSDmasses}
am_i\in\{0.00125\,,\ 0.00222\,,\ 0.005\,,\ 0.0075\,,\ 0.00889\}\ .
\end{equation}
All ensembles have the same bare coupling, and, in a mass-independent scheme,
the same lattice spacing \cite{GNS}.
We fitted the LSD data to LO dChPT in Ref.~\cite{GNS}.
Here, we repeat some of those fits replacing $\cl_d$ by $\cl_\D$,
keeping $\D$ as a free parameter.
Our results are shown in Table~\ref{tabD}.
These fits correspond to four fits presented
in Ref.~\cite{GNS}:  Fits \ref{tabD}A and
\ref{tabD}B are to be compared to the fits shown in Table~1 of Ref.~\cite{GNS},
while fits \ref{tabD}C and \ref{tabD}D
are to be compared with the third column of Table~3 and the second column
of Table~4 in Ref.~\cite{GNS}.

As discussed in great detail in Ref.~\cite{GNS}, it is not possible to fit all
parameters in the taste-breaking sector with the available LSD data.
Here we kept those taste-breaking parameters that gave rise to the
best fits of Ref.~\cite{GNS}.   Furthermore, in Ref.~\cite{GNS} we argued that
four-ensemble fits, which exclude the ensemble with the largest fermion mass,
are better behaved.  While the five-ensemble fits reported in Table~\ref{tabD}
already have good $p$-values, again we find that $p$-values for the
four-ensemble fits are significantly better.

Parameter values for $\g_*$ and $\log{d_0}$ are in good agreement with
the corresponding fits in Ref.~\cite{GNS}.
The parameters $d_1$ and $d_3$ are very poorly determined by the fits;
especially by those with four ensembles.    This is no surprise, as
$d_1$ and $d_3$ relate directly to the dilaton potential $\cl_\D$,
in which now a new parameter, $\D$, has been introduced.
The results for the taste-breaking parameters are in reasonable agreement
with Ref.~\cite{GNS} for the five-ensemble fit, and in good agreement for the
four-ensemble fit.   By holding $\D$ fixed in the fit, we verified that
in the limit $\D\to 4$ the results of Ref.~\cite{GNS} are reproduced.

The parameter $\D$ itself is reasonably well determined by each fit.
However, there is a visible difference between the four-ensemble
and five-ensemble fits.
From the four-ensemble fits, we conclude that $\D=3.5(7)$.
This is consistent with the hypothesis that dChPT, which predicts $\D\to 4$,
is the correct low-energy EFT.  The linear $\s$-model value, $\D=2$,
is disfavored.  By contrast, the values found in the five-ensemble fits
average to 2.8(7).  This is $1.7\s$ away from $\D\to 4$, and, in fact,
between the two options, it slightly favors the linear $\s$-model value.

%%%%%%%%%%%%%%%%%%%%%%%%%%%%
\subsection{\label{KMIDelta} The KMI data}
%%%%%%%%%%%%%%%%%%%%%%%%%%%%
We next turn to fits of the KMI data,
with $\cl_\D$ replacing $\cl_d$.   We first consider again window fits similar
to those of Table~\ref{KMIwindow}, but now with $\D$ an additional
free parameter.  The results are reported in Table~\ref{LatKMIDeltawindow}.
The fits are reasonably consistent with $\D=4$, while
the other parameters are generally consistent between
Tables~\ref{LatKMIDeltawindow} and \ref{KMIwindow}.
As before,  a constant $\g_m$ is
not sufficient to describe the KMI data over the full mass range.
However, while $\g_*$ varies with the mass range selected in the fit, $\D$ does not.
If we compare the values of $\D$ between two of the fits in
Table~\ref{LatKMIDeltawindow},
these values are always consistent within the smaller of the two errors
(with the exception of the second fit, for which $\D$ has an anomalously small
error).
The first and last values, $3.8(5)$ and $4.0(6)$,
coming from the lowest and highest mass ranges, are statistically
independent, in agreement with $\D=4$ and with each other.

%%%%%%%%%%%%%%%%%%%%%%%%%%%%%%%%%%%%%%%%%%%%%%%%%%%%%%%%%%%%%%%%%%%%%%%
\begin{table}[t]
%\begin{center}
\hspace{-0.1cm}
\begin{ruledtabular}
\begin{tabular}{|c|c|c|c|c|c|c|}
%\hline
range & 0.012--0.04 & 0.015--0.05 &0.02--0.06 & 0.03--0.07 & 0.04--0.08 & 0.05--0.1 \\
\hline
$\c^2$/dof & 9.16/5  & 8.11/5 & 4.81/5 & 3.42/5 & 3.69/5 & 3.82/5 \\
$p$-value  & 0.10 & 0.15 & 0.44 & 0.64 & 0.59 & 0.57  \\
\hline
$\D$ & 3.8(5) & 4.4(1) & 4.0(5) & 3.2(8) & 3.3(8) & 4.0(6) \\
$\g_*$  & 0.608(8) & 0.590(10) & 0.543(12) & 0.535(12) & 0.524(9) & 0.498(13) \\
$a\hf_\p$  & 0.008(7) & 0.0000(2) & 0.009(10) & 0.025(12) & 0.027(13) & 0.013(18) \\
$\td_1$ & 0.047(35) & 2(15) & 0.056(59) & 0.019(12)&0.019(12) &0.044(56)  \\
$-\log(ad_2)$  & 9.7(1.3) & 20(24) & 9.5(1.7) & 7.8(7) & 7.7(8) & 8.8(2.1)  \\
%\hline
\end{tabular}
\end{ruledtabular}
%\end{center}
%\vspace*{4ex}
\floatcaption{LatKMIDeltawindow}{{\it Fits of the KMI data to $\D$-dChPT
(with a constant $\g_m=\g_*$), with selections of five successive
fermion masses in Eq.~(\ref{barem}), shown in the top row.
}}
\end{table}
%%%%%%%%%%%%%%%%%%%%%%%%%%%%%%%%%%%%%%%%%%%%%%%%%%%%%%%%%%%%%%%%%%%%%%%%%

As in Sec.~\ref{KMI}, our next step is to consider fits to all, or most,
of the KMI data,
with $\cl_m$ of Eq.~(\ref{calBpionFfinal}), and a varying $\g_m$ as defined
in Eq.~(\ref{gammacub}).  As before, this introduces two more
parameters ($b$ and $c$) into the fits, for a total of seven parameters.
We will refer to this flavor of the low-energy lagrangian as $\g\D$-dChPT.

In Table~\ref{scanDelta} we show a scan in $\D$:
at each chosen value of $\D$, we fit the other six parameters.
The fit for $\D=3.9999$ coincides with fit~\ref{LatKMIgamma}A,
as one would expect. If we decrease $\D$, we find that the
$p$-value rapidly decreases, dipping below $0.01$ for $\D<3.8$.
We verified that the
$p$-value keeps decreasing  down to $\D=2$ (where the $p$-value
is of order $10^{-30}$).   If we increase $\D$ above 4,  the $p$-value
increases until $\D$ reaches  4.5,
where the $p$-value appears to start decreasing again.
However, we found that fits with $\D\ge 4.5$ become very difficult.
This is reflected in the very large errors in the six fit parameters:
for $\D=4.5$, essentially all of them are not determined by the fit.
We have repeated the fits of Table~\ref{scanDelta}
omitting the $am_0=0.1$ ensemble,
or the $am_0=0.1$ and $0.08$ ensembles, and we have also
redone such fits setting $c=0$ (as in fit~\ref{LatKMIgamma}D).
The conclusions are always the same as for the fits shown
in Table~\ref{scanDelta}.  The fit at $\D=3.9999$ is
consistent with the corresponding fit in Table~\ref{LatKMIgamma};
values of $\D$ below roughly $3.8$ are strongly disfavored; and the fit starts
to deteriorate at $\D=4.5$.
If we attempt to include $\D$ as a parameter in the fit itself
(instead of scanning over $\D$) fits appear to be unstable.

Given the difficulty fitting the KMI data with the $\cl_\D$ potential,
we have not attempted to include taste splittings in the KMI case.

%%%%%%%%%%%%%%%%%%%%%%%%%%%
%\newpage
\subsection{\label{disc} Discussion}
%%%%%%%%%%%%%%%%%%%%%%%%%%%
Taking the fits of the LSD and KMI data together,
it is clear that no very precise statement about the value of $\D$ can be made.
The KMI data appear to exclude the $\s$-model value $\D=2$.
dChPT, which corresponds to $\D\to 4$ with fixed $\g_m=\g_*$,
is consistent with the fits shown in Tables~\ref{tabD}
and \ref{LatKMIDeltawindow}.
An exception is the second window fit, fit~\ref{LatKMIDeltawindow}B,
which yields a result with a rather small error, $\D=4.4(1)$.
But clearly, this result does not account for the variation of $\D$
across all fits shown in Tables~\ref{tabD},
\ref{LatKMIDeltawindow} and \ref{scanDelta}.

Our results are consistent with those of Ref.~\cite{AIP3}.
The main difference is that the KMI data, which were not considered
in Ref.~\cite{AIP3}, present a much stronger lower bound on $\D$.

%%%%%%%%%%%%%%%%%%%%%%%%%%%%%%%%%%%%%%%%%%%%%%%%%%%%%%%%%%%%%%%%%%%%%%%%%%%%%%%%
\begin{table}[!t]
%\begin{center}
\hspace{-0.8cm}
\begin{ruledtabular}
\begin{tabular}{|l|l|c|c|c|l|c|c|c|}
%\hline
$\D$ & $\c^2$ & $p$-value & $a\hf_\p$   & $\td_1$ & $-\log(ad_2)$  & $\g_0$  & $b$ & $c$\\
\hline
4.5    & 11.0$^*$ & 0.69 & 0.00004(10) & 2.2(2.9)  & 19(17)              & 0.2(8.0) & $-$0.3(2.1)& $-$0.03(12) \\
4.4    & 10.2     & 0.75 & 0.00159(31) & 0.237(23) & 15(1)               & 2.33(79) & 0.73(41) & 0.07(5) \\
4.3    & 12.0     & 0.61 & 0.00384(39) & 0.123(6)  & 12.6(5)             & 2.04(49) & 0.81(32) & 0.11(5) \\
4.2    & 14.4     & 0.42 & 0.00613(42) & 0.084(3)  & 11.4(3)             & 1.88(35) & 0.87(28) & 0.14(5) \\
4.1    & 17.3     & 0.24 & 0.00831(43) & 0.063(2)  & 10.6(2)             & 1.78(28) & 0.93(25) & 0.17(5) \\
3.9999 & 20.7     & 0.11 & 0.01034(43) & 0.051(1)  & 10.1(2)             & 1.69(23) & 0.97(22) & 0.20(5) \\
3.9    & 24.7     & 0.04 & 0.01222(42) & 0.0423(7) & $\phantom{1}$9.7(1) & 1.62(20) & 1.00(21) & 0.23(5) \\
3.8    & 29.3     & 0.01 & 0.01396(41) & 0.0363(5) & $\phantom{1}$9.4(1) & 1.56(17) & 1.03(19) & 0.26(5) \\
%\hline
\end{tabular}
\end{ruledtabular}
%\end{center}
%\vspace*{4ex}
\floatcaption{scanDelta}{\it Fits of $M_\p^2/F_\p^2$ and $aF_\p$
to $\g\D$-dChPT, for fixed values of $\D$.
All fits have 14 degrees of freedom.
The fit with the asterisk may not have fully converged,
and its $\c^2$ value is an upper bound to the true minimum.
}
\end{table}
%%%%%%%%%%%%%%%%%%%%%%%%%%%%%%%%%%%%%%%%%%%%%%%%%%%%%%%%%%%%%%%%%%%%%%%%%%%%%%%%

As we show in App.~\ref{power counting}, for values of $\D$ not close to 4, no
power counting exists for the low-energy theory with $\cl_d$ replaced by
$\cl_\D$ of Eq.~(\ref{VDelta}).   However,
we do not wish to imply that attempts to understand data in terms
of models are not interesting.  Fits to models, including $\D$-dChPT
(with $\D$ not constrained to be close to 4),
can provide a valuable ``stress test'' of dChPT.
This is why we considered fits of the LSD and KMI data to $\D$-dChPT;
Ref.~\cite{AIP3} can be seen as a similar exploration of only the LSD data.

Fits of the LSD data, comparing in particular the values $\D=2$ and
$\D\to 4$, were considered also in Ref.~\cite{Kutietal}.\footnote{
  See also Ref.~\cite{Kutietal20} for related studies of the SU(3) theory
  with two sextet fermions, which also has a light flavor-singlet scalar.
  We recall, however, that dChPT is strictly speaking not applicable
  to this theory, as the Veneziano limit can be taken only
  for fermions in the fundamental representation.
}
There, it was found that both dChPT and $\D$-dChPT with $\D=2$
provide good fits to data using all five of the LSD
ensembles.   This finding agrees with our fits in Table~\ref{tabD}: fits
\ref{tabD}A and \ref{tabD}C are consistent with $\D=2$, but are less
than $\sim 2\s$ away from $\D=4$.

In summary, a precise determination of the favored value of $\D$
is not possible with presently available data.
Taking the results based on fits to both the LSD and KMI data together,
we arrive at an estimated range for $\D$,
\begin{equation}
\label{rangeD}
3.5<\D<4.5\ .
\end{equation}
Our lower bound is based on the four-ensemble fits to the LSD data,
which favor a value around $\D\sim 3.5$, combined with the $\g\D$-dChPT
scan of Table~\ref{scanDelta}, which strongly disfavors values below 3.8.
Any fit of the KMI data set must somehow account for the running of $\g_m$.
Including higher orders systematically is not an option here,
because, as we prove in App.~\ref{power counting}, the claim of Ref.~\cite{AIP3}
that $\D$-dChPT admits a systematic expansion is incorrect.
The model alternatives are to use a fixed value of $\g_m$
while limiting the mass range as in the ``window'' fits,
or else to use an explicitly varying $\g_m$ function.
As for the window fits, Table~\ref{LatKMIDeltawindow}
shows that $\D$ is rather insensitive to the mass range in the fit.
Also, while both the five-ensemble fits to the LSD data,
and some of the window fits to the KMI data allow
for $\D<3.5$, the fits of Table~\ref{scanDelta} to the KMI data
strongly disfavor $\D<3.8$.
Based on all fits together, the $\s$-model value $\D=2$ appears to be excluded.
Once again, the caveats discussed in the previous section
regarding the LSD and KMI data sets, and, in particular,
the lack of information about scaling violations,
apply also to our conclusions in this section.

%%%%%%%%%%%%%%%%%%%%%%%%%%%
%\newpage
\section{\label{conclusion} Conclusion}
%%%%%%%%%%%%%%%%%%%%%%%%%%%
Our main goal in this paper was to confront the EFT framework provided by dChPT
with the KMI data for the eight-flavor SU(3) gauge theory \cite{LatKMI}.
The KMI simulations were performed at larger fermion masses than
the LSD ones \cite{LSD2},
taking the theory further away from conformality.
Hence, even with the successful application of LO dChPT
to the LSD data, which we reported on in Ref.~\cite{GNS},
there is no guarantee
that LO dChPT can also be applied to the KMI data.

Indeed, we found that the full fermion-mass range of the KMI data
cannot be fitted to LO dChPT.  The natural next step would be
to attempt an NLO fit in dChPT.  However, as we explained in Sec.~\ref{KMI},
this is not feasible with presently available data.
First, the large number of parameters involved in any NLO dChPT fit
requires extensive precision data for a successful fit.  Moreover,
the KMI data set (and, likewise, the LSD data set)
has only a single lattice spacing,
making a continuum extrapolation impossible.

Instead, we introduced $\g$-dChPT, a model extension
of LO dChPT with a scale-dependent mass anomalous dimension,
which can be interpreted as arising from partially resumming higher orders
in the EFT expansion.  We found that $\g$-dChPT
provides a successful description of the KMI data over the entire mass range.

Given the success in describing the LSD data
using LO dChPT \cite{GNS}, and the KMI data using $\g$-dChPT with
a relatively simple {\em ansatz} for the $\g_m$ function, the question arises
whether $\g$-dChPT can be used to fit the LSD and KMI data simultaneously.
Over the KMI mass range, $\g_m$ would then have to increase as the fermion mass
is decreased, eventually saturating  to a constant when reaching
the lower LSD mass range (see Fig.~\ref{gammaplot}).
Once again, however, the inability to take the continuum limit
makes it impossible to carry out this program at this time.
The lack of information
on the lattice spacing dependence is even more severe when trying to
consider the LSD and KMI data sets together, because they were produced
with different lattice actions, and thus, their scaling violations
for any given physical observable are different functions of the corresponding
lattice spacing.

We also considered $\D$-dChPT---another generalization of LO dChPT
in which the dilaton potential is replaced by
a class of potentials depending on a new parameter $\D$.
We emphasize that $\D$-dChPT does not allow for a systematic power counting,
and should thus be considered a model, except in the limit $\D\to 4$
where dChPT is recovered.  $\D$-dChPT was
applied to the LSD data before \cite{AIP3}, where it was found that it is
difficult to determine the parameter $\D$ from these data.   We confirmed this
result, but found that the KMI data allow us to better constrain the value
of $\D$.  We used both the ``window'' fits in which $\D$-dChPT is applied
to subsets of the KMI ensembles, as well as a combination of
the two extensions of LO dChPT, with the $\D$ class of dilaton potentials
together with a varying $\g_m$.
We concluded that the preferred range of our combined analysis of the
LSD and KMI data is $3.5<\D<4.5$.  This is centered around $\D=4$,
where $\D$-dChPT reduces to LO dChPT.

Recently, LO dChPT has also been successfully applied to the light sector of
the SU(3) gauge theory with four light and six heavy flavors \cite{LSD10}.
dChPT provides for a systematic treatment of the pNGBs, the pions and the
dilaton, of a near-conformal gauge theory,
but it does rest on certain assumptions \cite{PP,largemass}.
These initial successes are thus encouraging.  We hope that, in the future,
more extensive and refined data will become available, allowing for
further and more stringent tests of dChPT.

%%%%%%%%%%%%%%%%%%%%%%%%%%%
\vspace{3ex}
%\newpage
\noindent {\bf Acknowledgments}
\vspace{2ex}
%%%%%%%%%%%%%%%%%%%%%%%%%%%

We thank Julius Kuti for discussions, and for asking probing questions
about the relation of $\g_m$ to dChPT.
MG's work is supported by the U.S. Department of
Energy, Office of Science, Office of High Energy Physics, under Award
Number DE-SC0013682.
YS is supported by the Israel Science Foundation
under grant no.~491/17.

%%%%%%%%%%%%%%%%%%%%%%%%%%%
%\newpage
\appendix
\section{\label{massindep} Scale setting prescription}
%%%%%%%%%%%%%%%%%%%%%%%%%%%
Any analysis of lattice data requires a scale setting prescription,
and the basic choice is between mass-independent or mass-dependent
prescriptions.  In this paper, as in Ref.~\cite{GNS}, we opted for
a scale-independent prescription, and confirmed the self-consistency
of this choice by checking that the values of $a\hB_\p$ on all ensembles
agree within error (see Sec.~\ref{KMI}).

Here we discuss the alternative of using a mass-dependent prescription.
In QCD simulations it is common nowadays to use the gradient flow scale $t_0$
for scale setting \cite{MLflow}.  In particular, the ensemble value of $t_0$
can be used for a mass-dependent prescription.
What makes $t_0$ particularly convenient for setting the scale is
that it can be determined with high precision, and it admits
a chiral expansion, with non-analytic terms in the quark mass
entering only at NNLO \cite{BG}.  By contrast, as we showed in Ref.~\cite{GNS},
in dChPT there is no (useful) chiral expansion for $t_0$.
This implies that one cannot derive expansions for dimensionless quantities
such as $\sqrt{t_0}M_\p$, $\sqrt{t_0}F_\p$, \etc, using dChPT.

If we are interested in dChPT fits, we are thus unable to use $t_0$
for scale setting.   Instead, we may consider using a physical
quantity such as $F_\p$ for a mass-dependent scale setting prescription.
As explained in Sec.~\ref{hadrons},
with our mass-independent scale setting, the basic fit has six parameters,
two of which, namely $d_2$ and $\hf_\p$, have mass dimension equal to one.
On each ensemble, we fit $aF_\p$ and $M_\p^2/F_\p^2$ to Eqs.~(\ref{quantb})
and~(\ref{MFratio}), respectively.  In addition, we treat the fermion mass $am$
as a data point with a small fictitious error, in order to determine
the expectation value $v$ of the dilaton field on each ensemble,
via Eq.~(\ref{sprew}).

If we use $F_\p$ for mass-dependent scale setting,
we may still fit $M_\p^2/F_\p^2$ to Eq.~(\ref{MFratio}) as before.
In addition, combining Eqs.~(\ref{quantb}) and~(\ref{sprew}) together,
we may fit $m/F_\p$ as
\begin{equation}
\label{mFpi}
  \frac{m}{F_\p} = \frac{d_2}{\hf_\p}\,\frac{1}{\td_1(3-\g_m)}\,v\,e^{F(v)}\ ,
\end{equation}
and use it to determine $v$, now as a function of $m/F_\p$.
This procedure gives us access only to the ratio $d_2/\hf_\p$,
instead of to $ad_2$ and $a\hf_\p$ separately, as in the fitting
procedure of Sec.~\ref{hadrons}.

We tried to repeat the fits from Table~\ref{LatKMIgamma},
using Eqs.~(\ref{MFratio}) and~(\ref{mFpi}).
The result was that almost all fit parameters remained completely undetermined.
The nominal fit quality was always very high ($p$-value $\ge 0.97$),
consistent with the failure of the fits to resolve the parameters.

We believe the problem is caused by the much smaller
number of degrees of freedoms which are available for the
mass-dependent fitting procedure.  As an example,
consider fit~\ref{LatKMIgamma}A, which uses all 10 ensembles,
and has $2\times 10 - 6 = 14$ degrees of freedom.
As noted in Sec.~\ref{varying}, since the $v_i$'s are determined
in terms of the $am_i$'s using Eq.~(\ref{sprew}), this does not change
the number of degrees of freedom.
By contrast, within the mass-dependent fitting procedure,
we have in total only 20 relations to determine both
the 5 fit parameters and the 10 auxiliary $v_i$'s.
This leaves us with just 5 degrees of freedom,
which apparently is just not enough to resolve the fit parameters.

Although we were unable to actually perform a fit with
a mass-dependent prescription for scale setting, we may consider the following
``thought experiment.''  Assume that the data allowed for fits
with a mass-dependent prescription, and that the results of those fits
are in agreement with Table~\ref{LatKMIgamma}.
This would mean, in particular,
that the value of the new dimensionless fit parameter, $d_2/\hf_\p$,
obtained from fitting Eq.~(\ref{mFpi}) is consistent with
the results for $a\hf_\p$ and $ad_2$ reported in Table~\ref{LatKMIgamma}.
Now, while we determine all parameters in the large-mass regime,
which is where both the LSD and KMI data are, $\hf_\p$ and $d_2$
are LECs that characterize the massless theory.
As we discussed in Sec.~\ref{KMIdisc}, the values of $a\hf_\p$ and $ad_2$
extracted from the LSD and the KMI data sets appear to be in conflict,
both with the chiral limit values of $t_0$ (determined from
a phenomenological fit), and with each other.
It is interesting to check what is the situation for
the dimensionless ratio $d_2/\hf_\p$.
Comparing the value of this ratio
using the results in Table~\ref{LatKMIgamma} to those from Ref.~\cite{GNS}
reveals that there is still a significant conflict,
of roughly the same size as for $a\hf_\p$, though smaller than for $ad_2$.
We conclude that restricting ourselves to a mass-dependent
scale setting prescription would not by itself alleviate
the problem of the long extrapolation from the large-mass regime
to the chiral limit.

%%%%%%%%%%%%%%%%%%%%%%%%%%%
%\newpage
\section{\label{power counting} Power counting}
%%%%%%%%%%%%%%%%%%%%%%%%%%%
In Ref.~\cite{AIP3} it was proposed that $\D$-dChPT---in which
the potential $\cl_d$ of Eq.~(\ref{hatpota}) is replaced
by $\cl_\D$ of Eq.~(\ref{VDelta}), with $\D$ a new free parameter---admits
a systematic power counting.
In this appendix, we show that the arguments given
in Ref.~\cite{AIP3} are not correct.

The potential~(\ref{VDelta}) was already considered in Refs.~\cite{GGS,CM}.
In those papers it was assumed that the lagrangian of
the underlying theory contains an operator with scaling dimension
$\D$, with some unspecified value of $\D$, and a coupling which may be small.
This naturally leads to the consideration of potentials such as Eq.~(\ref{VDelta})
in the EFT describing the same theory at low energy.

By contrast, here the underlying theory is known: it is the
asymptotically free SU(3) gauge theory with $N_f=8$ Dirac fermions
in the fundamental representation.  This theory does not fall into
the class of theories considered in Refs.~\cite{GGS,CM}.

It is instructive to briefly recall how the breaking of scale invariance
is introduced into the (massless) quantum theory;
and then, how this breaking translates to the EFT \cite{PP}.
As can be seen in Eq.~(\ref{dimreg}), regularizing the bare lagrangian
requires the introduction of a scale factor, $\m_0^{d-4}$,
with the limit $d-4\to 0$ to be taken after renormalization.
By letting $\m_0$ transform according to Eq.~(\ref{bareb}),
we may promote $\m_0$ to a spurion, formally restoring scale invariance.

In making the transition to the EFT, we will want to use the well-known fact
that the EFT lagrangian must be analytic in the spurion fields,
if the underlying lagrangian is analytic in the (same set of) spurions.
Correlation functions can then be generated by differentiating
the partition function of the EFT with respect to the spurion fields,
and compared with their counterparts in the underlying theory by
applying the same derivatives again.  This matching procedure
fixes the LECs of the EFT order by order, according to
the power counting.

A technical obstacle is that the action~(\ref{dimreg}) is
non-analytic in the spurion $\m_0$.  To overcome this problem,
we introduce a new spurion field $\s(x)$, and replace
\begin{equation}
\label{sigma}
\m_0 \Rightarrow \hm_0 e^{\s(x)}\ .
\end{equation}
The new scale transformation rules replacing Eq.~(\ref{bareb}) are
\begin{subequations}
\label{strans}
\begin{eqnarray}
\label{stransa}
  \s(x) &\to& \s(\l x)+\log\l \ ,
\\
\label{stransb}
  \hm_0 &\to& \hm_0 \ .
\end{eqnarray}
\end{subequations}
Now $\hm_0$ is invariant under a scale transformation, which in turn
is ``carried'' by the constant mode of the new spurion field.
Writing $\s(x) = \s_0 +\d\s(x)$,
with the constraint $\int d^dx\, \d\s(x)=0$, it follows that
\begin{equation}
\label{s0transf}
\s_0\to\s_0+\log\l\ .
\end{equation}
With this replacement, the bare action~(\ref{dimreg}) becomes
\begin{equation}
\label{dimregs0}
S= \hm_0^{d-4}\int d^dx\, e^{(d-4)\s(x)}\cl(x)\ .
\end{equation}
Classically, the $\s(x)$ dependence vanishes for $d\to 4$,
showing that any dependence of the renormalized theory on $\s(x)$
represents quantum breaking of scale invariance \cite{CDJ,PP,gammay}.
Since the underlying theory is now analytic in the spurion field $\s(x)$,
so must be the EFT \cite{PP}.  Note that if, instead, one were to use $\m_0$
as a scale spurion, there would be no reason for the EFT to be analytic
in $\m_0$, for the simple reason that the underlying theory~(\ref{dimreg}) is
non-analytic in $\m_0$.\footnote{%
  The same statement applies if the constant spurion $\m_0$ is
  promoted to a field.
}

In Ref.~\cite{AIP3}, the starting point of the argument was to assume that
the lagrangian of the low-energy theory depends analytically
on a spurion field $\tm(x)$, with the scale transformation rule\footnote{%
  The spurion $\tm(x)$ is denoted as $\l(x)$ in Ref.~\cite{AIP3},
  see Eq.~(A1) therein.  We have reserved $\l$ for the
  scale transformations parameter, which in turn is denoted
  as $e^\r$ in Ref.~\cite{AIP3}.
}
\begin{equation}
\label{AIPtrans}
  \tm(x) \to \l^{4-\D} \tm(\l x) \ .
\end{equation}
It is clear from the previous discussion that the underlying gauge theory
does not accommodate such a spurion.  Comparing transformation rules,
one can, however, make the identification
\begin{equation}
\label{matchAIP}
  \tm(x) \equiv e^{(4-\D)\s(x)} \ .
\end{equation}
As we have just shown, the correct EFT must be analytic in $\s(x)$,
but not in $e^{\s(x)}$ (nor in any power of $e^{\s(x)}$).
It follows immediately that the EFT must be analytic in $\log\tm(x)$,
but not in $\tm(x)$ itself.
This proves that the arguments of Ref.~\cite{AIP3}
are not valid, because the incorrect assumption that the EFT is analytic
in $\tm(x)$ served as their starting point.

While this proves that the power counting claimed in Ref.~\cite{AIP3}
is unfounded, several comments are in order.

First, we draw the reader's attention that in Sec.~\ref{vary}
we made use of the original spurion $\m_0$, instead of $\s(x)$.
The reason is that our goal in Sec.~\ref{vary} was to derive
the extension of LO dChPT to the case of a running $\g_m$.
This requires mainly the consideration of renormalization-group
and scale transformation properties, and, for this purpose,
using $\m_0$ as a (constant) scale spurion is sufficient.\footnote{%
  In accordance with our general reasoning, in Eq.~(\ref{calBpion}),
  $\cl_m$ indeed depends on $\log\m_0$.
}
The $\g$-dChPT framework developed in Sec.~\ref{vary} does deviate from the
strict power counting of dChPT \cite{PP}, though it can be viewed as
a resummation of contributions from all orders under the
assumption that these dominate.
As for establishing the power counting itself, this necessitates
the replacement of $\m_0$ by $\hm_0 e^{\s(x)}$, \seef\ Eq.~(\ref{sigma}).
Correspondingly, the transformation rules~(\ref{strans}) take over
the transformation rule of $\m_0$ in Eqs.~(\ref{bare}) and~(\ref{transeff}).
For the actual proof of the power counting, and a detailed discussion of
the assumptions that it requires, we refer to Refs.~\cite{PP,largemass}.

A key step in constructing a power-counting scheme is the identification,
in the underlying theory, of a small parameter in terms of which
the EFT expansion is to be organized.
In ordinary ChPT, the small parameter is the fermion mass $m$, which is
also the ``expectation value'' of the chiral spurion, $\svev{\c(x)}=m$.
Chiral symmetry is restored for $m\to 0$, which in turn allows
to establish that the pion mass is parametrically small.

By contrast, in Ref.~\cite{PP}, the small parameter controlling the hard breaking
of scale invariance was identified as $n_f-n_f^*$, serving as a proxy
for the $\b$ function at the chiral symmetry breaking scale.
More precisely, the hypothesis made in Ref.~\cite{PP} is
\begin{equation}
  \tb \sim |n_f-n_f^*|^\h \ , \qquad n_f \nearrow n_f^* \ ,
\label{tbg}
\end{equation}
for some $\eta>0$, where
\begin{equation}
  \tb = \frac{\m}{4\a} \frac{\partial \alpha}{\partial\mu} \ ,
\label{betatilalph}
\end{equation}
and $\a$ is the 't Hooft coupling, $\a=g^2 N/(4\p)$,
evaluated at the chiral symmetry breaking scale.
While we often assume $\h=1$ for simplicity, including earlier in this paper,
this assumption is not essential.
The power counting is valid for any fixed $\h>0$; the $\h$ dependence
is restored trivially via the
substitution $|n_f-n_f^*| \Rightarrow |n_f-n_f^*|^\h$.

The small parameter $|n_f-n_f^*|^\h$ does not appear explicitly
in the underlying lagrangian,
and, in particular, it is not identified with the expectation value of $\s(x)$.
Indeed, unlike chiral symmetry, which is restored for $m=0$,
there is no fixed value of $\s(x)$ for which scale invariance
is {\it not} broken.  Rather, the expansion of correlation functions
in powers of $\s(x)$ corresponds to an expansion in the number of insertions
of the trace anomaly.  In the massless limit, every such insertion
is proportional to the $\b$ function at the chiral symmetry breaking scale,
hence to $|n_f-n_f^*|^\h$.
For this argument to work, it is crucial to use the spurion field $\s(x)$,
and not $\m_0$ or $\tm(x)$.
The role of $\s(x)$, or of its constant mode $\s_0$,
is analogous to that of the $\theta$ parameter in the large-$N_c$ limit
of ChPT in which the U(1)$_A$ symmetry is restored.
For a detailed comparison, we refer to Ref.~\cite{PP}.
The upshot is that one cannot establish a relation between
the expectation value of $\m_0$ or $\tm(x)$
and the $\b$ function of the underlying theory.
Hence, even if one were to allow the low-energy theory
to depend only on integer powers of the $\tm(x)$ spurion,
as was postulated in Ref.~\cite{AIP3}, there is no reason to assume
that its expectation value should tend to zero
when the conformal window is approached.
Both assumptions of
Ref.~\cite{AIP3}, analyticity in the spurion $\tm$, and its smallness in the
conformal limit, are thus in conflict with the properties
of the underlying theory, and, in general, not valid.

This concludes our discussion of the claims made in Ref.~\cite{AIP3} with regard
to power counting.   But, a little more can be said about the connection
of the potential $\cl_\D$ with dChPT, which corresponds to the limit $\D\to 4$.
According to the dChPT power counting developed in Ref.~\cite{PP},
the scale invariant dilaton potential $e^{4\t}$ is multiplied by
a potential $\tV_d(\t)$ that breaks scale invariance,
\begin{equation}
\label{Vdgen}
 \tV_d(\t) = \sum_{n=0}^\infty \frac{\tc_n}{n!}\, \t^n \ .
\end{equation}
The LECs $\tc_n$ scale as $\tc_n \sim |n_f-n_f^*|^{n\h}$,
and with the power counting (compare Eq.~(\ref{pc}))
\begin{equation}
\label{psq}\
  p^2 \sim m \sim |n_f-n_f^*|^\h \sim 1/N \ ,
\end{equation}
it follows that the term $\frac{\tc_n}{n!} \t^n$ can only appear
at N$^{n-1}$LO in dChPT.  In particular,
the tree-level potential $V_d(\t)$ of Eq.~(\ref{Vd})
obtained after the $\t$ shift corresponds to $c_1=\tc_1=-4\tc_0$.

When $\D$ is close to 4, we may identify $\tV_d(\t)=V_\D(\t)$,
which, using Eq.~(\ref{VDeltab}), implies that $\tc_0=-c_1/\D$ and
\begin{equation}
\label{VDexpand}
\tc_n = (4c_1/\D) (\D-4)^{n-1} \ ,  \qquad n\ge 1 \ .
\end{equation}
The first two terms in the expansion reproduce
the LO potential $V_d(\t)$.\footnote{%
  Except for the innocuous replacement $c_1 \Rightarrow 4c_1/\D$.
}
It follows that, for any fixed value of $\D$
such that $|\D-4| \sim |n_f-n_f^*|^\h$ with any $\h>0$,
the $V_\D(\t)$ potential will inherit the power counting of dChPT.
The same is not true for values of $\D$ not close to 4,
and thus, it is also not true for the low-energy lagrangian
in which $\D$ is treated as a free parameter.

%\clearpage
%\newpage
\vspace{3ex}
%%%%%%%%%%%%%%%%%%%%%%%%%%%

\end{document}